\documentclass[amsmath,amssymb,amsbsy,prb,twocolumn,superscriptaddress,showpacs]{revtex4-1}

\usepackage{graphicx}
\usepackage{rotating}
\usepackage{dcolumn}
\usepackage{bm}
\usepackage{mathrsfs}
\usepackage{color}
\usepackage{leftidx}
\usepackage[breaklinks,colorlinks = true,linkcolor = red,urlcolor=cyan,citecolor=red]{hyperref}
\usepackage{varwidth}

\newcommand{\avg}[1]{\left< #1 \right>} 

\begin{document}

\title{Three-dimensional nematic spin liquid in the stacked triangular lattice 6H-B structure}

\author{Kyusung Hwang}
\affiliation{Department of Physics and Centre for Quantum Materials, University of Toronto, Toronto, Ontario M5S 1A7, Canada}

\author{Tyler Dodds}
\affiliation{Department of Physics and Centre for Quantum Materials, University of Toronto, Toronto, Ontario M5S 1A7, Canada}

\author{Subhro Bhattacharjee}
\affiliation{Department of Physics and Centre for Quantum Materials, University of Toronto, Toronto, Ontario M5S 1A7, Canada}
\affiliation{Department of Physics and Astronomy, McMaster University, Hamilton, Ontario L8S 4M1, Canada}

\author{Yong Baek Kim}
\affiliation{Department of Physics and Centre for Quantum Materials, University of Toronto, Toronto, Ontario M5S 1A7, Canada}
\affiliation{School of Physics, Korea Institute for Advanced Study, Seoul 130-722, Korea}


\date{\today}

\begin{abstract}

Recently, a number of experiments indicate the possible presence of spin liquid phases  in quantum magnets with spin-1/2 and spin-1 moments sitting on
triangular-lattice-based structures {in Ba$_3$CuSb$_2$O$_9$ and Ba$_3$NiSb$_2$O$_9$ respectively}. In relation to {these experiments}, several theoretical proposals have been made  for spin liquid phases and spin-liquid-like
behaviors on the stacked triangular lattice.
While the crystal structures of
these materials are currently under debate, it is nonetheless interesting to
understand possible spin liquid phases on such frustrated lattices. In this
work, we apply Schwinger boson mean-field theory and projective symmetry group
(PSG) analysis to investigate spin liquid phases on the fully three-dimensional
6H-B structure, in contrast to previous works that considered two-dimensional
systems. We find that a nematic $Z_2$ spin liquid phase, where the
lattice-rotational symmetry is spontaneously broken, is the most promising spin
liquid phase that is consistent with spiral magnetic ordering in the classical
limit. We discuss the implications of our results to future theoretical and
experimental works.

\end{abstract}

\maketitle

\section{INTRODUCTION}

Spin-$1/2$ moments on two or three dimensional lattices may not order, even at zero temperature, owing to competing interactions and/or quantum fluctuations and form a highly entangled state of quantum matter.\cite{1973_anderson,2002_wen,2006_kitaev,2008_plee, QSL_Balents, 2011_yan,2012_jiang_1,2012_jiang}  Such a correlated state, called a quantum spin liquid (QSL),\cite{1973_anderson,2002_wen,2006_kitaev,2008_plee, QSL_Balents} is capable of supporting quasiparticle excitations that carry only a fraction of the quantum numbers of the underlying electrons that make up the spin system.  

This possibility of emergent quantum number fractionalization and other exotic
excitations like {\it artificial} photons,\cite{2002_wen}  resulting from
strong many-body correlation effects in condensed matter systems, have fuelled
massive interest in quantum spin liquids both theoretically and experimentally.
A large number of Mott insulators in various frustrated lattice geometries have
been investigated, and several candidate materials have been identified.
\cite{2008_plee,QSL_Balents} Most of these materials, such as
$\kappa$-(BEDT-TTF)$_2$Cu$_2$(CN)$_3$,\cite{2003_shimizu}
Cs$_2$CuCl$_4$,\cite{2003_coldea}
EtMe$_3$Sb[Pd(dmit)$_2$]$_2$,\cite{2011_yamashita} (spin-1/2 on triangular
lattice) and Herbertsmithite (ZnCu$_3$(OH)$_6$Cl$_2$, spin-1/2 kagome
lattice)\cite{Herbertsmithite_Mendels,Herbertsmithite_Helton,Herbertsmithite_Han} are two dimensional.
However,  some three dimensional systems, like the hyper-kagome lattice
Na$_4$Ir$_3$O$_8$,\cite{2007_okamoto} several candidate quantum spin-ice
materials on the pyrochlore lattices,\cite{2011_ross} and more recently the
stacked compounds
Ba$_3$CuSb$_2$O$_9$\cite{Zhou_Ba3CuSb2O9_2011,Nakatsuji_Ba3CuSb2O9_2012,Quilliam_Ba3CuSb2O9_2012}
(S=1/2) and Ba$_3$NiSb$_2$O$_9$\cite{Cheng_Ba3NiSb2O9_2011} (S=1) have shown
much promise. In {the} light of the above developments, it is useful to explore and
understand the behavior of quantum spin liquids in associated frustrated three
dimensional lattices.

\begin{figure}
 \centering
 \includegraphics[width=0.9\linewidth]{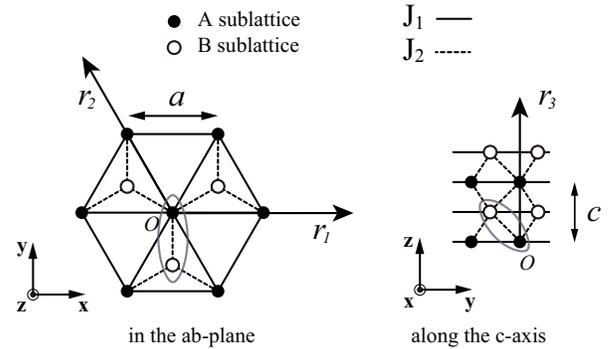}
 \caption{The 6H-B lattice structure and coordinate system used throughout the paper. 
 The 6H-B lattice consists of the AB-stacked triangular lattices in three dimensions.
 The left figure shows neighboring two sublattices projected into the $ab$-plane.
 For the B sublattice, the couplings with $J_1$ are omitted to simplify the figure.
 The right figure shows the lattice structure along the $c$-axis.
 The ellipse indicates the unit cell located at the origin of the coordinate system. ${\bf r}_1, {\bf r}_2, {\bf r}_3$ denotes the three axes of lattice translation.
 \label{fig:coordinate_system}}
\end{figure}

In this paper, we study bosonic spin liquids that can be realized in quantum
magnets on a three dimensional crystal structure built by stacking triangular
lattices in a staggered fashion. This so-called 6H-B phase is obtained in the
P$6_3/mmc$ lattice structure, as shown in Fig. \ref{fig:coordinate_system}. In
particular, we focus on a spin-$1/2$ antiferromagnetic Heisenberg model with
intra-layer and inter-layer spin exchanges.  The two interactions compete and
the model is frustrated when the intra and inter-layer exchanges are
comparable. While earlier studies {on the 2D honeycomb limit} found
interesting ground state degeneracies and non-collinear magnetic states
selected by quantum order-by disorder,\cite{Mulder_2010} our Schwinger boson
mean field theory (SBMFT) captures  a {\it 3D $Z_2$ spin liquid}, a {\it 3D
$U(1)$ spin liquid} and a layered, effectively {\it 2D, $Z_2$ spin liquid} in
addition to the usual magnetically ordered phases. Using projective symmetry
group (PSG) classification for the spin-liquids,\cite{2002_wen,2006_wang} we find the
above bosonic spin liquids that are consistent with various lattice symmetries.
A central finding of our SBMFT study is that a 3D {\it nematic} $Z_2$ spin
liquid, which spontaneously breaks lattice rotation symmetries, is found to be
energetically more stable than an isotropic $Z_2$ spin liquid in the
corresponding parameter regime.  Interestingly, we find that,
unlike the isotropic case, the nematic spin
liquid is naturally connected to the classical {magnetic orders}.
The breaking of lattice rotation symmetry in the nematic
spin liquid bears a characteristic signature in the two-spinon excitation
spectrum for the nematic spin liquid. Such {{two}}-spinon spectrum
is relevant to neutron scattering experiments. Within mean-field theory we find
several quantum phase transitions among spin-liquids, and between spin-liquids
and magnetically ordered phases. While some of them, like the transition
between the collinear {{N\'eel}} and the 3D $U(1)$ spin liquid, are
most likely rendered discontinuous by gauge {fluctuations} beyond
the mean field, others like the transition between the 3D, $Z_2$ and the spiral
are likely to remain continuous. The transition between various magnetically
ordered phases and the spin liquids can be understood in terms of condensation
of the spinons\cite{1994_chubukov, 2009_cenke} and a charge-2 Higgs field
(discussed later).

As noted earlier, a partial motivation for studying the above spin model in
this lattice structure stems from recent interests in two new candidate QSLs,
viz., Ba$_{3}$CuSb$_{2}$O$_{9}$ (spin-1/2) and (6H-B-)Ba$_{3}$NiSb$_{2}$O$_{9}$,
(spin-1) where initial experiments suggested a 6H-B structure.
\cite{Zhou_Ba3CuSb2O9_2011,Cheng_Ba3NiSb2O9_2011,Nakatsuji_Ba3CuSb2O9_2012}
These compounds do not show magnetic order down to a few hundred mK, despite 
Curie-Weiss temperatures, as measured from the high temperature susceptibility,
{(greater than 50 K)}.
Furthermore, they show a large
intermediate temperature window where the magnetic specific heat shows a linear
temperature dependence suggesting the possibility of an unconventional spin
state. While recent experiments on the spin-$1/2$ Cu compound revealed a
different crystal structure for
Ba$_{3}$CuSb$_{2}$O$_{9}$,\cite{Zhou_Ba3CuSb2O9_2011,Nakatsuji_Ba3CuSb2O9_2012,Nasu_spin_1/2_2012}
it may be still useful to think about possible spin-$1/2$ and/or spin-1 spin
liquids on the 6H-B phase in search of three dimensional spin liquids and
possible exotic quantum phase transitions.  {{This approach is in
contrast to the previous theoretical studies that mainly considered two
dimensional spin models for the
compounds.}}\cite{{Serbyn_spin_1_2011},{Xu_spin_1_2012},{Chen_spin_1_2012},{Bieri_spin_1_2012},{Thompson_spin_1/2_2012},{Corboz_spin_1/2_2012},{Nasu_spin_1/2_2012}}

The rest of this paper is structured as follows. In Sec. \ref{sec:structure},
we give an overview of the P6$_3/mmc$ lattice structure obtained in the 6H-B
phase, discuss its symmetries, and introduce a frustrated Heisenberg model.
Next, we discuss the Schwinger boson formalism in Sec. \ref{sec:SBMFT}, and
derive the mean-field Hamiltonian, which can describe the transition between
classical order and {as well as a} regime of stronger quantum fluctuations. To classify
distinct mean-field ans\"atze with a given symmetry, we introduce the PSG
analysis in Sec. \ref{sec:PSG-analysis}, and describe the allowed ansatz with
the full symmetry of the lattice, along with those that break lattice
rotational symmetry {spontaneously}. We work out the magnetic phases obtained in the classical
limit of our model in Sec. \ref{sec:semi-classical-approach}, and note that
this behavior cannot be fully captured from the symmetric spin liquid ansatz.
In fact, we find that the symmetric spin liquid is generally energetically
inferior to the nematic spin liquid in the relevant part of the {SBMFT} phase diagram.
The full phase diagram{s} as a function of $\kappa$ and $J_2/J_1$
for the different {ans\"atze} are discussed in Sec.
\ref{sec:mean-field-phase-diagram}. We also point out the nature of different
phase transitions and indicate the underlying mechanisms to study them. Finally,
the structure of the two-spinon excitations is discussed in Sec.
\ref{sec:two-spinon spectra}. We end by summarizing our results and discussing
their implications in Sec. \ref{sec:summary-outlook}. The details of the PSG
calculations, derivations of the ans\"atze {and other details} are {given} in different
{a}ppendices.

\section{Lattice structure and Spin Hamiltonian for the 6H-B Phase \label{sec:structure}}

The 6H-B structure may be thought of as a collection of triangular lattices
stacked along the $c$-axis, with two consecutive layers being offset as shown in
Fig. \ref{fig:coordinate_system}. Sites on even layers form the triangular A
sublattice. Those on odd layers form the triangular B sublattice, and are offset
above the center of the triangles of the A sublattice.

On such a lattice, we consider a spin-1/2 model where the predominant interactions between spins on the lattice are nearest-neighbor antiferromagnetic super-exchange interactions. Interactions within the A and B triangular planes have strength $J_1$, and those between neighboring planes have strength $J_2$. Both couplings may be comparable in the case of systems such as Ba$_3${Cu}Sb$_2$O$_9$, where the exchange paths are mediated through intermediate oxygen atoms. The resulting Heisenberg Hamiltonian is 
\begin{align}
		H = J_1 
		\hspace{-12pt}
		\sum_{\langle i,j \rangle  \textrm{ in plane}}
		\hspace{-12pt}
		{\bf S}_i \cdot {\bf S}_j
		+ J_2 
		\hspace{-22pt}
		\sum_{\langle i,j \rangle  \textrm{ between planes}} 
		\hspace{-23pt}
		{\bf S}_i \cdot {\bf S}_j,
		\label{eq:heisenberg-hamiltonian}
\end{align}
where $J_1$, $J_2>0$.

This model has two simple limits: (i) $J_1 \gg J_2$ and (ii) $J_1 \ll J_2$. In
the first case, the model is reduced to the two-dimensional triangular lattice
Heisenberg model. It is well known that the ground state of the model has
120$^\circ$ non-collinear (spiral) N\'{e}el order.\cite{1999_capriotti} In the
other limit, the system has a three-dimensional bipartite lattice structure
without frustration so that it allows collinear N\'{e}el order as its ground
state.\cite{2011_albuquerque} However, in the regime where both couplings
are comparable ($J_1 \approx J_2$), the interactions compete with each other,
and this may lead to a quantum spin liquid ground state. At this point, we also
note (as discussed in detail later) that in the classical limit, the above
spin model has the spiral ordered ground state in the intermediate region of
coupling constants (see Fig. \ref{fig:classical_phase_diagram}).

We proceed by noting various symmetries of the spin model in anticipation of
our future Projective Symmetry Group calculation. The lattice is described by
the following primitive vectors, as seen in Fig. \ref{fig:coordinate_system}:
\begin{align}
 {\bf R}_1 = a \hat{x},
 {\bf R}_2 = a \left( - \frac{1}{2} \hat{x} + \frac{\sqrt{3}}{2} \hat{y} \right),
 {\bf R}_3 = c \hat{z},
\label{eq:lattice vectors}
\end{align}
where $a$ and $c$ are the sublattice spacings in the $a$-$b$ plane and along the $c$-axis, respectively. We write lattice coordinates as ${\bf r}_A = \sum_{n=1}^{3}r_n{\bf R}_n \equiv (r_1,r_2,r_3)_A$ {for the A sublattice}, and ${\bf r}_B = {\bf r}_A + {\bf r}_{BA} {\equiv (r_1,r_2,r_3)_B}$ {for the B sublattice}, where ${\bf r}_{BA} = - {\bf R}_1 /3 - 2{\bf R}_2 /3 + {\bf R}_3 /2$. In this coordinate representation, $r_n~(n=1,2,3)$ is an integer. The system has the following symmetries: spin-rotation, time-reversal, and space group symmetries. The first two symmetries are discussed in the next sections. The spin model (\ref{eq:heisenberg-hamiltonian}) has the space group $P6_3/mmc$ with following seven generators:{\cite{space_group_table}}
\begin{itemize}
		\item Translations $T_1$ and $T_2$ within the triangular plane, and another, $T_3$, along the $c$-axis.
		\item A $120^\circ$-rotation $R$ around the $z$-axis, centered on an A site.
		\item A reflection $\Pi_1$ through the $y$-$z$ plane, and another, $\Pi_2$, through the $x$-$y$ plane.
		\item An inversion $\Xi$ through the midpoint between neighboring A and B sites.
\end{itemize}
Consequently, the space group transformations are defined as
\begin{subequations}
\label{eq:space-group-transformations}
\begin{eqnarray}
 T_1 
 & : &
 (r_1,r_2,r_3)_p \rightarrow (r_1+1,r_2,r_3)_p,
 \\
 T_2
 & : &
 (r_1,r_2,r_3)_p \rightarrow (r_1,r_2+1,r_3)_p,
 \\
 T_3
 & : &
 (r_1,r_2,r_3)_p \rightarrow (r_1,r_2,r_3+1)_p,
 \\
 R
 & : &
 (r_1,r_2,r_3)_p \rightarrow (-r_2+\delta_{p,B},r_1-r_2+\delta_{p,B},r_3)_p,
 \nonumber\\
 \\
 \Pi_{1}
 & : &
 (r_1,r_2,r_3)_p \rightarrow (-r_1+r_2,r_2,r_3)_p,
 \\
 \Pi_{2}
 & : &
 (r_1,r_2,r_3)_p \rightarrow (r_1,r_2,-r_3-\delta_{p,B})_p,
 \\
 \Xi
 & : &
 (r_1,r_2,r_3)_p \rightarrow (-r_1,-r_2,-r_3)_{\bar{p}},
\end{eqnarray}
\end{subequations}
where $p~(\bar{p})=A,B~(B,A)$ refers to the two sublattices.

Having determined the symmetries of our spin model, in the next section we begin our Schwinger boson mean-field analysis, and proceed to identify the various possible bosonic spin liquids within a projective representation of the above symmetry group.
\section{SCHWINGER BOSON MEAN-FIELD THEORY AND GAUGE STRUCTURE\label{sec:SBMFT}}

In the Schwinger boson mean-field theory,  \cite{SBMFT_Arovas,SBMFT_Sachdev_1,SBMFT_Sachdev_2} the spin operator at site $i$ is represented in terms of bosonic spinons $b_{i\mu}$:
\begin{align}
 S_i^a = \frac{1}{2} b_{i\mu}^{\dagger} \sigma_{\mu\nu}^a b_{i\nu},
\label{eq:Schwinger-boson-representation}
\end{align}
where $a=x,y,z$, $\mu,\nu=\uparrow,\downarrow$,  $\{ \sigma^a \}$ are the Pauli matrices, and a sum over repeated Greek indices is assumed hereafter. In this formalism, operators of any spin quantum number $S$ can be represented, which is given by the on-site density constraint,
\begin{align}
 \kappa=b_{i\mu}^{\dagger} b_{i\mu}=2S.
 \label{eq:constraint-on-the-spin-moment-size}
\end{align}
Within the mean-field theory, the constraint is implemented on average. {In general, $\kappa/2$ represents the spin quantum number and  acts as a parameter that determines the degree to which quantum fluctuations are important to our Hamiltonian.} The $\kappa~(\sim S)\rightarrow \infty$ limit corresponds to classical limit.\cite{SBMFT_Sachdev_2}

Following the usual SBMFT techniques,\cite{SBMFT_Sachdev_1,SBMFT_Sachdev_2} we now consider the mean-field decoupling of the Heisenberg terms. With the help of the constraint (\ref{eq:constraint-on-the-spin-moment-size}), we can rewrite these terms as
\begin{align}
 {\bf S}_i \cdot {\bf S}_j 
 =
 -\frac{1}{2} \hat{\eta}_{ij}^{\dagger} \hat{\eta}_{ij} + \frac{\kappa^2}{4},
 \label{eq:MF_decoupling}
\end{align}
where $\hat{\eta}_{ij} = b_{i\mu} \epsilon_{\mu \nu} b_{j\nu}$ and
$\epsilon_{\mu \nu}$ is totally antisymmetric tensor with $\epsilon_{\uparrow
\downarrow}=1$. In this form, the mean-field decoupling is straightforward. 
After the mean-field decoupling, we employ a Lagrange multiplier, $\lambda_i$, for each site
to implement the constraint (\ref{eq:constraint-on-the-spin-moment-size}).
In the resulting quadratic mean-field Hamiltonian,
we introduce the parameter $x_{i\mu} = \avg{b_{i\mu}}$ which represents the
bosonic condensate fraction:
\begin{equation}
 b_{i\mu} \rightarrow x_{i\mu} + b_{i\mu}.
\end{equation}
Now $b_{i\mu}$ implies the noncondensate part of the original bosonic spinon.
If $x_{i\mu}\neq0$, 
there is condensation of the spinons, and hence a long range magnetically ordered state
with spontaneously broken spin-rotation symmetry.
However, if the
spinons are gapped then there is no condensate, and the spin-rotation
symmetry is preserved. This is a spin liquid state with gapped bosonic spin-1/2
(spinon) excitations. In three dimensions, it can be
either a 
$Z_2$ or U(1) spin liquid. In the former case, in addition to the spinons, there is an
emergent gapped non-magnetic excitation called the {\it
vison}.\cite{1999_senthil} On the other hand, in the U(1) spin liquid, there is
an emergent gapless {\it photon} (with two polarization modes) and a gapped
magnetic monopole excitation.\cite{2002_wen} These issues are discussed in some
detail later.  The mean-field Hamiltonian is written as
\begin{eqnarray}
 H_{MF} 
 &=&
 \sum_{i>j} \left( -\frac{1}{2} J_{ij} \eta_{ij} \right) b_{i\mu}^{\dagger} \epsilon_{\mu \nu} b_{j\nu}^{\dagger} + \textup{H.c.}
 \nonumber\\
 &+&
 \sum_{i>j} \left( -\frac{1}{2} J_{ij} \eta_{ij} \right) \left( x_{i\mu}^* \epsilon_{\mu \nu} x_{j\nu}^* \right) + \textup{c.c.},
 \nonumber\\
 &+&
 \sum_{i} \lambda_i \left( b_{i\mu}^{\dagger} b_{i\mu} + |x_{i\mu}|^2- \kappa \right)
 \nonumber\\
 &+&
 \sum_{i>j} \left( \frac{1}{2} J_{ij} \left| \eta_{ij} \right|^2 + \frac{1}{4} J_{ij} \kappa^2 \right)
 \label{eq:mean-field-hamiltonian-final}
\end{eqnarray}
where $\eta_{ij} = x_{i\mu} \epsilon_{\mu \nu} x_{j\nu} + \langle b_{i\mu} \epsilon_{\mu \nu} b_{j\nu} \rangle$ 
is the mean-field expectation value of the bond parameter. 
The above mean-field Hamiltonian is then solved self-consistently using the following saddle-point equations:
\begin{align}
 \frac{\partial \left<H_{MF}\right>}{\partial {\eta}_{ij}^*} = 0,
 ~~
 \frac{\partial \left<H_{MF}\right>}{\partial {x}_{i\mu}^*} = 0,
 ~~
 \frac{\partial \left<H_{MF}\right>}{\partial {\lambda}_i} = 0.
 \label{eq:self_consistent_equations}
\end{align}

There exists a $U(1)$ gauge redundancy in the Schwinger boson representation of the spin operator (\ref{eq:Schwinger-boson-representation}). That is, under the transformation
\begin{align}
b_{i\mu} \to e^{i\phi_i} b_{i\mu},
\end{align}
the spin operator ${\bf S}_i$ in (\ref{eq:Schwinger-boson-representation}) is invariant. However, the mean-field {parameter} transforms as
\begin{align}
\eta_{ij} \to e^{-i(\phi_i+\phi_j)} \eta_{ij}.
\label{eq:gauge-transf-MF}
\end{align}
For non-bipartite lattices, when a particular mean-field state is chosen such that $\eta_{ij} \neq 0$ is fixed, the above $U(1)$ gauge invariance is broken down to $Z_2$, since now the mean-field Hamiltonian is gauge invariant only for $\phi_i = 0,\pi$.\cite{2002_wen,2006_wang} The resulting state is a $Z_2$ spin liquid. The gauge redundancy described above, however, suggests that different $Z_2$ spin liquid {ans\"atze} may be connected by gauge transformations and hence correspond to the same physical state.

In the next section, we provide the classification scheme for physically distinct $Z_2$ spin liquid phases, using a PSG analysis.\cite{2002_wen,2006_wang}

\section{PROJECTIVE SYMMETRY GROUP ANALYSIS FOR THE MEAN-FIELD ANS\"ATZE\label{sec:PSG-analysis}}

\subsection{Brief Overview}

At the mean-field level, different spin liquids are characterized by different PSGs, which are projective extensions of the symmetry group of the Hamiltonian.\cite{2002_wen,2006_wang,1989_hammermesh}

For a given spin liquid ansatz (in our case a choice of $\eta_{ij}$), the PSG
is the set of operations $\{G_X X\}$, where $X$ is an element of the symmetry
group and $G_X$ is the associated gauge transformation, that {leave}
the mean-field Hamiltonian invariant. Here, we will describe the general
framework to determine $\{G_X\}$ for a particular symmetry group (SG). The
details of the calculations are given {in} Appendix
\ref{appendix:PSG-construction} and \ref{appendix:PSG-construction-2}. Note
that this analysis characterizes spin-liquid states without long-range order;
that is, mean-field Hamiltonians (\ref{eq:mean-field-hamiltonian-final}) in the
absence of a condensate $x_{i\mu}$. Considering the case of $Z_2$ spin
liquids, we define \begin{align}
		h_{ij} \equiv \eta_{ij}b^{\dag}_{i\mu} \epsilon_{\mu\nu} b^{\dag}_{j\nu},
\end{align}
and find that under the combined effect of a symmetry transformation, $X$, and associated gauge transformation $G_X$, we have
\begin{align}
		(G_X X) h_{ij} = \eta_{ij} e^{-i\left(\phi_X[X(i)] + \phi_X[X(j)]\right)}
	 b^{\dag}_{X(i)\mu} \epsilon_{\mu\nu} b^{\dag}_{X(j)\nu}.
	 \label{eq:action-of-G_S-S}
\end{align}
Hence, for $\{G_X X\}$ that leave $H_{MF}$ invariant, symmetry-related mean-field parameters are found by
\begin{align}
		\eta_{X(i)X(j)} = \eta_{ij} e^{-i\left(\phi_X[X(i)] + \phi_X[X(j)]\right)}.
		\label{eq:symmetry-related-bonds}
\end{align}

As mentioned in Sec. \ref{sec:SBMFT}, gauge transformations generally change
$H_{MF}$, but a particular subset, called the Invariant Gauge Group (IGG) of
the PSG, may leave it invariant. The IGG is a projective extension of the
identity operation of the symmetry group. One can choose a gauge in which the
IGG is independent of the site.\cite{2002_wen} For $Z_2$ spin liquids,
{IGG=$\{+1,-1\}$}. The structure of the IGG determines the nature of the
low-energy gauge fluctuations around $H_{MF}$. For $Z_2$ spin liquids, these
fluctuations are gapped. There is a gapped non-magnetic vortex-like excitation
that carries the flux of the $Z_2$ gauge field. 
However, for $U(1)$
spin liquids, there is an emergent gapless photon, as well as a gapped magnetic
monopole excitation, the {latter} resulting from the compactness of
the U(1) gauge group.\cite{2002_wen} 

We note that if $G_X X$ leaves $H_{MF}$ invariant, so does $W G_X X$ for $W
\in$ IGG, which means that a class of $G_X$ is defined only up to elements of
the IGG. To generate restrictions for $\{G_X\}$, we note that products of
$G_{X_a} X_a$ that are physically equivalent to an identity transformation must
be in the IGG.\cite{2002_wen,2006_wang} Since each operation leaves $H_{MF}$
invariant, the total transformation has no net physical transformation and also
leaves $H_{MF}$ invariant. Multiplication rules among the generators of the
symmetry group generate precisely these products of transformations. We write
these rules in the form $X_a X_b X_c^{-1} X_d^{-1} = I$, and note that upon
inclusion of the corresponding gauge transformations, we have \begin{align}
	(G_{X_a} X_a)
	(G_{X_b} X_b)
	(G_{X_c} X_c)^{-1}
	(G_{X_d} X_d)^{-1}
	\in \mathrm{IGG}.
	\label{eq:identity-to-igg}
\end{align}

Motivated by the classical solution of the Heisenberg model  (\ref{eq:heisenberg-hamiltonian}) {(as determined in Sec. \ref{sec:PSG-analysis}),} we consider two sets of symmetry groups.
\begin{subequations}
\begin{eqnarray}
 &&
 {\textup{SG}_1}=\textup{Span}\{ T_1, T_2, T_3, \Pi_1, \Pi_2, \Xi \},
 \\
 &&
 {\textup{SG}_2}=\textup{Span}\{ T_1, T_2, T_3, \Pi_1, \Pi_2, \Xi, R \}.
\end{eqnarray}
\end{subequations}
The group $\textup{SG}_2$ consists of the full set of lattice symmetries given in (\ref{eq:space-group-transformations}). However, we will find that the resulting ansatz with $\textup{SG}_2$ is too restrictive to be consistent with the classical limit of this model. We find that it does not lead to the right magnetic ordered state in the classical limit. Thus, we also consider the symmetry group $\textup{SG}_1 (\subset \textup{SG}_2)$  by removing the rotation $R$, which will be shown to provide a spin liquid phase consistent with the spiral magnetic order in the classical limit. For each of these symmetry groups, we find the gauge transformations {$\{ G_X \}$} associated with the symmetry operations, and generate the resulting ans\"atze $H_{MF}$ using Eq. (\ref{eq:symmetry-related-bonds}).

\subsection{$Z_2$ Spin Liquid Ans\"atze}

In this subsection, we discuss the ans\"atze for both symmetry groups $\textup{SG}_1$ and $\textup{SG}_2$. A brief sketch of their derivation is given in Appendix \ref{appendix:SG2_ansatze}.
{{Mean-field Hamiltonians for the ans\"atze are provided in Appendix}} \ref{appendix:mean-field Hamiltonians}.

\begin{figure}
 \centering
 \includegraphics[width=0.5\linewidth]{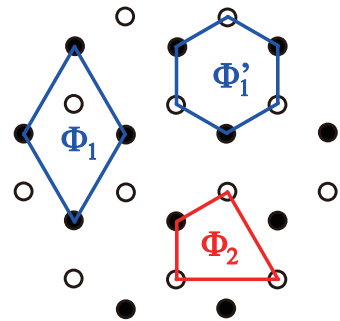}
  \caption{
  Gauge fluxes to differentiate the ans\"atze of SG$_1$ and the loops where they are defined.
  The gauge flux on each loop is defined with (\ref{eq:gauge-flux}).
  The values of the fluxes for the ans\"atze are listed in Table \ref{tab:ansatze of SG_1}.
  }
 \label{fig:gauge_flux_loop}
\end{figure}

We find four different ans\"atze for $\textup{SG}_1$ and only one ansatz for $\textup{SG}_2$.
The ans\"{a}tze are differentiated by the gauge fluxes of $\{ \eta_{ij} \}$ through various loops with even length.
The gauge flux $\Phi$ on a loop with length $2n$ is defined through
\begin{eqnarray}
 &&
 \eta_{i_1 i_2} (-\eta_{i_2 i_3}^*) \cdots \eta_{i_{2n-1} i_{2n}} (-\eta_{i_{2n} i_1}^*) 
 \nonumber\\
 &=&
 | \eta_{i_1 i_2} | | \eta_{i_2 i_3} | \cdots | \eta_{i_{2n-1} i_{2n}} | | \eta_{i_{2n} i_1} |  
 \cdot
 e^{i \Phi}.
 \label{eq:gauge-flux}
\end{eqnarray}
Under the effects of a gauge transformation, (\ref{eq:gauge-transf-MF}), 
this flux is invariant, 
providing a clear way to distinguish among the ans\"atze that have the same symmetry group.\cite{Tchernyshyov_flux_2006} 
To differentiate the ans\"atze,
we consider the gauge fluxes on the three kinds of loops shown in Fig. \ref{fig:gauge_flux_loop}.
Filled and empty circles denote neighboring A and B sublattices, viewed along $c$-axis.
The flux $\Phi_1$ is defined in any rhombus on each triangular sublattice layer.,
${\Phi}'_1$ is the flux defined in any hexagonal loop between neighboring sublattice layers.
The last flux is $\Phi_2$ on a kite-shaped loop.
For the ans\"atze of SG$_1$, 
the loop consists four different mean-field links
$\{ \eta_{1\alpha}, \eta_{1\beta}, \eta_{2\alpha}, \eta_{2\beta} \}$,
which will be defined below.
The three fluxes for the ans\"atze of SG$_1$ are listed in Table \ref{tab:ansatze of SG_1}.
Interestingly, $\Phi_1={\Phi}'_1$ for the four ans\"atze of SG$_1$.
The ans\"atze are named with the gauge-invariant fluxes ($\Phi_1$,$\Phi_2$).

\begin{table}[b]
\caption{
Classification of the mean-field ans\"atze of SG$_1$.
$\Phi_1$, ${\Phi}'_1$, $\Phi_2$ are defined on the closed loops in Fig. \ref{fig:gauge_flux_loop}. 
$n_1~(=0 ~ \textup{or} ~ 1)$ is an integer variable characterizing the projective symmetry group of SG$_1$ [see (\ref{eq:PSG_for_SG_1})].
$m~(=0 ~ \textup{or} ~ 1)$ is an integer variable reflecting the phase difference among the four mean-field parameters, $\eta_{1\alpha}$, $\eta_{1\beta}$, $\eta_{2\alpha}$, $\eta_{2\beta}$ [defined in (\ref{eq:four-indep-MF-parameters-2})].
$n_s$ is the number of sites in a unit cell. 
${\bf R}'_2={\bf R}_1+2{\bf R}_2$.
The ($0,0$)-flux and ($0,\pi$)-flux ans\"atze are translationally invariant,
whereas the ($\pi,0$)-flux and ($\pi,\pi$)-flux ans\"atze are not invariant.
The latter ans\"atze have doubled unit cell with four sites.
The (0,$\pi$)-flux ansatz for $\textup{SG}_1$ includes the ansatz for $\textup{SG}_2$ 
as a special case with $\eta_{1\alpha} = \eta_{1\beta}$ and $\eta_{2\alpha} = \eta_{2\beta}$.
\label{tab:ansatze of SG_1}
}
\begin{ruledtabular}
\begin{tabular}{lcccccc}
 ansatz & $\Phi_1={\Phi}'_1$ & $\Phi_2$ & $n_1$ & $m$ & $n_s$ & lattice vectors
 \\
 \hline
 $(0,0)$-flux & 0 & 0 & 0 & 1 & 2 & $\{ {\bf R}_1, {\bf R}_2, {\bf R}_3 \}$
 \\
 $(0,\pi)$-flux & 0 & $\pi$ & 0 & 0 & 2 & $\{ {\bf R}_1, {\bf R}_2, {\bf R}_3 \}$
 \\
 \hline
 $(\pi,0)$-flux & $\pi$ & 0 & 1 & 0 & 4 & $\{ {\bf R}_1, {\bf R}'_2, {\bf R}_3 \}$
 \\
 $(\pi,\pi)$-flux & $\pi$ & $\pi$ & 1 & 1 & 4 & $\{ {\bf R}_1, {\bf R}'_2, {\bf R}_3 \}$
 \\
\end{tabular}
\end{ruledtabular}
\end{table}

\paragraph{SG$_1$:}

\begin{figure}
 \centering
 \includegraphics[width=1.0\linewidth]{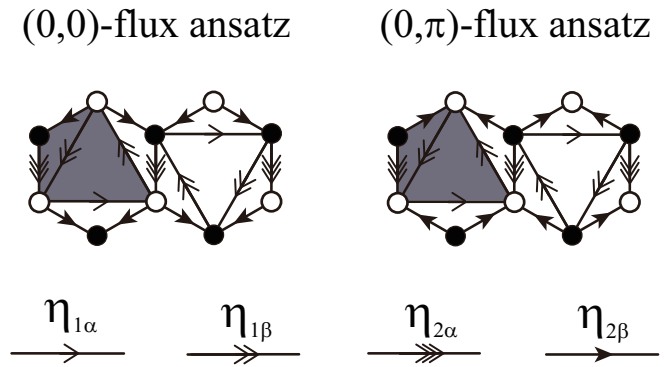}
 \caption{
 ($0,0$)-flux ansatz and ($0,\pi$)-flux ansatz of the symmetry group SG$_1$.
  The figure shows the directions and magnitudes of the allowed mean-field parameters, viewed along the $c$-axis. 
  Filled and empty circles denote A and B sublattices, respectively. 
  Arrows indicate the directions of the four allowed positive, real mean-field parameters $\eta_{1\alpha}$, $\eta_{1\beta}$, $\eta_{2\alpha}$, and $\eta_{2\beta}$. 
  The shaded region indicates the inside of the three-dimensional path around which the gauge flux $\Phi_2$ is defined.
 The flux $\Phi_2$ distinguishes the two ans\"atze.
  }
\label{fig:ansatze-definitions}
\end{figure}

As mentioned above, four different ans\"atze are found for the symmetry group $\textup{SG}_1$:
the ($0,0$)-flux ansatz, ($0$,$\pi$)-flux ansatz, ($\pi,0$)-flux ansatz, and ($\pi$,$\pi$)-flux ansatz.

The ($0,0$)-flux ansatz and ($0$,$\pi$)-flux ansatz are depicted in Fig. \ref{fig:ansatze-definitions}. 
Both ans\"atze are translationally invariant.
To explicitly preserve time-reversal symmetry, 
we work in a gauge where these ans\"atze have real-valued mean-field parameters $\eta_{ij}$. 
Since $\eta_{ji} = - \eta_{ij}$, we denote the directions of the positive parameters in Fig. \ref{fig:ansatze-definitions}. 
In the symmetry group SG$_1$, due to the lack of rotational symmetry, 
there are two different in-plane mean-field parameter magnitudes $\eta_{1\alpha}$ and $\eta_{1\beta}$, and two inter-plane ones $\eta_{2\alpha}$ and $\eta_{2\beta}$.

The (0,0)- and (0,$\pi$)-flux ans\"atze are distinguished by the flux $\Phi_2$.
The shaded region in Fig. \ref{fig:ansatze-definitions} indicates the inside of the loop where $\Phi_2$ is defined.
In fact, the flux $\Phi_2$ is defined only when every $\eta_{ij}$ along the loop is non-zero. In the mean-field theory, we will come across two special cases with $\eta_{1\alpha} = \eta_{1\beta} \equiv \eta_1 $ and $\eta_{2\alpha} = \eta_{2\beta} \equiv \eta_2$: one where $\eta_1 = 0$, and the other where $\eta_2 = 0$. In the first case, $\eta_1=0$, our (0,0)- and (0,$\pi$)-flux ans\"atze become gauge-equivalent. Furthermore, the connectivity changes so that the two sublattices become bipartite. As we show later, in this case we actually have a three-dimensional gapped $U(1)$ spin liquid, which is in principle a stable phase, unlike in two dimensions. We denote this as the {3D-$U(1)$} state. In the second case, $\eta_2=0$, our two ans\"atze are identical, giving a two-dimensional $Z_2$ spin liquid with zero flux through rhombus plaquettes in the triangular planes ($\Phi_1=0$). We denote this as the {2D-$Z_2$} state.
It is one of the two $Z_2$ spin liquids allowed {with} the symmetry of the 2D triangular lattice.\cite{2006_wang} The symmetry group of the 6H-B structure does not include the 2D triangular lattice symmetry group as a subgroup, due to the lack of the six-fold rotational symmetry. As a result, the PSG analysis on the 6H-B structure, in the 2D limit, recovers only the $Z_2$ spin liquid with zero flux within the rhombus plaquettes in the triangular planes. 
The other $Z_2$ spin liquid with $\pi$ flux of the 2D triangular lattice appears as a saddle point, not the energy minimum, 
in the $J_2=0$ limit of ($\pi$,0)- and ($\pi$,$\pi$)-flux ans\"atze.
For details, readers are referred to Appendix \ref{appendix:(pi,zero/pi)_solution}.

Before moving on to the case of SG$_2$,
we briefly mention about the ($\pi,0$)-flux and ($\pi$,$\pi$)-flux ans\"atze.
Those ans\"atze are not {manifestly} invariant under translation.
They break the translation symmetry along $r_2$-direction so that they have doubled unit cell with four sites.
The mean-field configurations of the ans\"atze are shown in Fig. \ref{fig:pi-zero_ansatz}.

\paragraph{SG$_2$:}

The PSG analysis of the full symmetry group $\textup{SG}_2$ finds only one allowed ansatz, which has $\Phi_1={\Phi}'_1=0$ and $\Phi_2=\pi$. This is related to the (0,$\pi$)-flux ansatz with $\textup{SG}_1$, for the case where the mean-field magnitudes acquire rotational symmetry, {\it i.e.}, $\eta_{1\alpha} = \eta_{1\beta}$ and $\eta_{2\alpha} = \eta_{2\beta}$. Thus, the (0,$\pi$)-flux ansatz for $\textup{SG}_1$ includes the ansatz for $\textup{SG}_2$ as a special case.

Having discussed the allowed {ans\"atze} for our symmetry groups
$\textup{SG}_1$ and $\textup{SG}_2$, we consider the phases obtained in
different parameter regimes.
\section{CLASSICAL ORDERING\label{sec:semi-classical-approach}}

We begin by studying the classical ground state of the spin model
(\ref{eq:heisenberg-hamiltonian}), in the context of Schwinger boson mean-field
theory. The classical limit is obtained by taking the $\kappa \to \infty$ limit
of the SBMFT. From the scaling behavior of the mean-field solution for $\kappa
\gg 1$, ($\eta_{ij} \sim \kappa$, $x_{i\mu} \sim \sqrt{\kappa}$, $\lambda \sim
\kappa$) the ground state energy can be written as 
\cite{SBMFT_Sachdev_2}
\begin{eqnarray}
 E_c 
 =
 \frac{\langle H_{MF} \rangle}{\kappa^2}
 &=&
 \sum_{i>j} \left( \frac{J_{ij}}{2} \left| \tilde{\eta}_{ij} \right|^2 + \frac{J_{ij}}{4} \right)
 \nonumber\\
 &+&
 \sum_{i>j} \left( -\frac{J_{ij}}{2} \tilde{\eta}_{ij} \right) \left( \tilde{x}_{i\mu}^* \epsilon_{\mu \nu} \tilde{x}_{j\nu}^* \right) + \textup{c.c.}
 \nonumber\\
 &+&
 \sum_{i} \tilde{\lambda}_i \left( \left| \tilde{x}_{i\mu} \right|^2 - 1 \right),
 \label{eq:epsilon_c}
\end{eqnarray}
In the above expressions,
$\tilde{\eta}_{ij} = \eta_{ij} / \kappa$, $\tilde{x}_{i\mu} = x_{i\mu} / \sqrt{\kappa} $, 
$\tilde{\lambda} = \lambda / \kappa $.
The ground state is determined by solving the following mean-field equations:
\begin{align}
 \frac{\partial E_{c}}{\partial \tilde{\eta}_{ij}^*} = 0,
 ~~
 \frac{\partial E_{c}}{\partial \tilde{x}_{i\mu}^*} = 0,
 ~~
 \frac{\partial E_{c}}{\partial \tilde{\lambda}_i} = 0.
 \label{eq:semiclassical-self-consistent-equations}
\end{align}
Making use of the above mean-field equations, we rewrite the ground state energy as
\begin{align}
 E_{c}
 =
 \sum_{i>j} \frac{J_{ij}}{4} {\bf S}_i^c \cdot {\bf S}_j^c + \sum_{i} \tilde{\lambda}_i \left( |{\bf S}_i^c| - 1 \right),
 \label{eq:classical_Heisenberg_model}
\end{align}
where the classical spin vector is given by
\begin{align}
 {\bf S}_i^c = \tilde{x}_{i\mu}^{*} \boldsymbol{\sigma}_{\mu \nu} \tilde{x}_{i\nu}.
 \label{eq:classical_spin}
\end{align}
This is precisely the classical Heisenberg model with the constraint of normalized spin vectors.
The solution of (\ref{eq:classical_Heisenberg_model}) can be obtained via the Luttinger-Tisza method\cite{Luttinger_tisza_1,Luttinger_tisza_2,Luttinger_tisza_3} as follows:
\begin{subequations}
\label{eq:classical-solution}
\begin{eqnarray}
 &&
 {\bf S}_A^c ({\bf r}) 
 =
 {\bf n}_1 \textup{cos} [{\bf Q}_c \cdot {\bf r}]
 +
 {\bf n}_2 \textup{sin} [{\bf Q}_c \cdot {\bf r}],
 \\
 &&
 {\bf S}_B^c ({\bf r}) 
 =
 - {\bf n}_1 \textup{cos} [{\bf Q}_c \cdot {\bf r} - \Theta({\bf Q}_c)]
 - {\bf n}_2 \textup{sin} [{\bf Q}_c \cdot {\bf r} - \Theta({\bf Q}_c)],
 \nonumber\\
\end{eqnarray}
\end{subequations}
where the unit vectors ${\bf n}_1$ and ${\bf n}_2$ 
satisfy ${\bf n}_1 \cdot {\bf n}_2 = 0$.
The ordering wave vector, ${\bf Q}_c$, and the relative phase, $\Theta({\bf Q}_c)$, are defined by 
the following equations:
\begin{subequations}
\begin{eqnarray}
 &&
 3 
 + 2 \textup{cos}[{\bf Q}_c \cdot {\bf R}_1] 
 + 2 \textup{cos}[{\bf Q}_c \cdot {\bf R}_2] 
 + 2 \textup{cos}[{\bf Q}_c \cdot ({\bf R}_1 + {\bf R}_2)]
 \nonumber\\
 &&
 =
 ( {J_2}/{J_1} )^2,
 \\
 &&
 ({J_2}/{J_1}) e^{i \Theta({\bf Q}_c)}
 =
 1
 +
 e^{i{\bf Q}_c \cdot {{{\bf R}_2}}}
 +
 e^{i{\bf Q}_c \cdot ( {\bf R}_1 + {\bf R}_2 ) }.
\end{eqnarray}
\label{eq:Q_c_theta_Q_c}
\end{subequations}
The equations (\ref{eq:Q_c_theta_Q_c}) are for the cases of $J_2/J_1 \leq 3$. When $J_2/J_1 > 3$, the ground state solution at $J_2/J_1 = 3$ is given for the entire range.

\begin{figure}
 \includegraphics[width=0.8\linewidth]{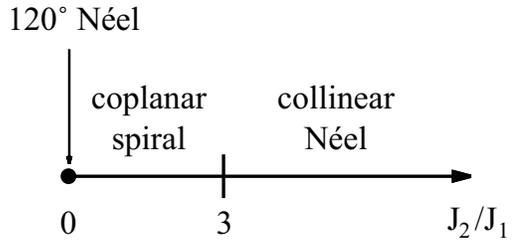}
 \caption{Phase diagram of classical magnetic order of the Hamiltonian (\ref{eq:heisenberg-hamiltonian})
 as a function of $J_2/J_1$.
 \label{fig:classical_phase_diagram}
 }
\end{figure}

\begin{figure}
 \centering
 \includegraphics[angle=270,width=1.0\linewidth]{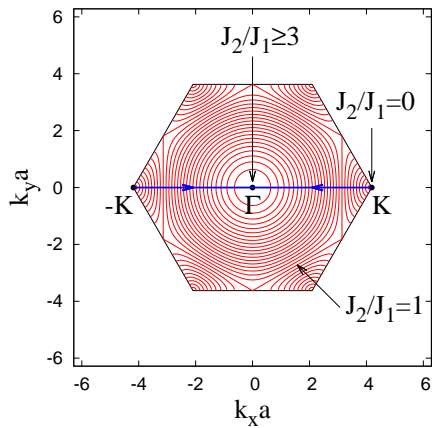}
 \caption{(Color online) Magnetic ordering wave vectors (${\bf Q}_c$) of the classical solutions. The wave vectors for the degenerate states at a given $J_2/J_1$ are plotted with a red line in the momentum space, and are determined by (\ref{eq:Q_c_theta_Q_c}). The degeneracy is lifted by the quantum order by disorder effect. We show the selected states within the (0,0)-flux ansatz with a blue line, which are described in Sec. \ref{sec:mean-field-phase-diagram}. The black hexagon denotes the first Brillouin zone at $k_z=0$. \label{fig:Q_vector}}
\end{figure}

Figure \ref{fig:classical_phase_diagram} shows the phase diagram obtained by solving the equations (\ref{eq:Q_c_theta_Q_c}). In all phases, ${\bf Q}_c \cdot \hat{z} = 0$. When $J_2 = 0$, the triangular layers are decoupled and each plane has 120$^{\circ}$ N\'{e}el order with ${\bf Q}_c$=$\pm$K=$\pm(\frac{4\pi}{3a},0,0)$ in the first Brillouin zone. Conversely, when $J_2/J_1 \geq 3$, ${\bf Q}_c = {\bf 0}$, and there is inter-plane collinear N\'eel order. In the intermediate case, $ 0 \leq J_2 / J_1 \leq 3$, the degenerate spiral ordering wavevectors interpolate between these behaviors, and are shown in Fig. \ref{fig:Q_vector}.

Throughout our classical Luttinger-Tisza calculations, we find that ${\bf Q}_c \cdot \hat{z} = 0$. Hence, for the classical order, we can consider the system as comprised of two layers, which have a honeycomb lattice structure when looked at along the $c$-axis. Thus, our classical solution is equivalent with those of the $J_1$-$J_2$ Heisenberg model on the honeycomb lattice.\cite{Mulder_2010,Chen_spin_1_2012}

The classical spiral states for $0 < J_2/J_1 < 3$ spontaneously break lattice rotational symmetry $R$, as well as spin-rotational symmetry. In the next section, Sec. \ref{sec:mean-field-phase-diagram}, we will see that the (0,$\pi$)-flux ansatz with the full symmetry of the lattice does not host a condensate that gives any such spiral order in the classical limit. In contrast, the (0,0)-flux ansatz does so, and selects the particular states shown in Fig. \ref{fig:Q_vector} {{(blue line)}}. 

With the classical limit understood, we now  present the phase diagram through the parameter range of quantum fluctuations $\kappa$.
\section{MEAN-FIELD PHASE DIAGRAM\label{sec:mean-field-phase-diagram}}

In this section, we study the phase diagram for the different ans\"atze of
symmetry groups SG$_1$ and SG$_2$, as a function of $\kappa$ and $J_2/J_1$,
where $\kappa \to \infty$ is the classical limit.
As mentioned in previous section,
we focus on the (0,0)-flux and (0,$\pi$)-flux ans\"atze among the four ans\"atze of SG$_1$.
The ($\pi$,0)-flux and ($\pi$,$\pi$)-flux ans\"atze are discussed  
in Appendix \ref{appendix:(pi,zero/pi)_solution}, since
they are always higher mean-field energies, and more importantly
they do not reproduce the correct magnetic order
in the semi-classical limit.



\subsection{Phase Diagrams for SG$_1$}

\subsubsection{(0,0)-flux Ansatz\label{sec:0-flux}}

\begin{figure}[ht]
 \centering
 \includegraphics[width=0.9\linewidth]{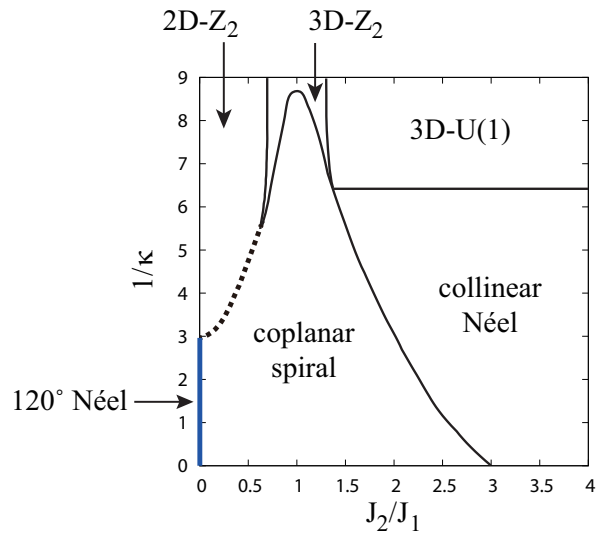}
 \caption{(Color online) Mean-field phase diagram of the (0,0)-flux ansatz. 
 In the phase diagram, there are three long range orders for large $\kappa$, 
 which are the 120$^\circ$ N\'eel, coplanar spiral, and collinear N\'eel states, 
 and three spin liquid phases for small $\kappa$,
 which are the $2D$-$Z_2$, $3D$-$Z_2$, and $3D$-$U(1)$ states.
 The thick blue line indicates the region where the 120$^\circ$ N\'eel state appears.
 This mean-field phase diagram recovers the classical phases in Fig. \ref{fig:classical_phase_diagram} in large $\kappa$ limit.
 At the boundary between the coplanar spiral and $2D$-$Z_2$ states (denoted with the dotted line), 
 the phase transition is discontinuous, while a transition at any other phase boundary is continuous. \label{fig:0-flux-phase-diagram}}
\end{figure}

\begin{figure}
 \centering
 \includegraphics[angle=270,width=0.8\linewidth]{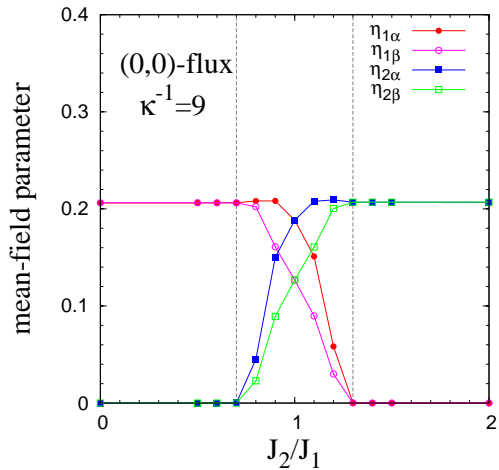}
 \caption{
(Color online)
 Mean-field parameters of the (0,0)-flux spin liquid states at $\kappa^{-1}=9$.
 In the $2D$-$Z_2$ spin liquid state ($J_2/J_1 \leq 0.7$),
 $\eta_{1\alpha}=\eta_{1\beta} \ne 0$ and $\eta_{2\alpha}=\eta_{2\beta}=0$.
 The opposite is true for the $3D$-$U(1)$ spin liquid state ($J_2/J_1 \geq 1.3$).
 In the $3D$-$Z_2$ spin liquid state ($0.7 < J_2/J_1 < 1.3$),
 every mean-field parameter is nonzero.
 These features in the mean-field parameters indicate that  
 the $3D$-$Z_2$ spin liquid state breaks the 120$^\circ$ lattice-rotation symmetry,
 while the rotation symmetry is preserved in the other two states.
 \label{fig:MF_solutions_SL}}
\end{figure}

\begin{figure}
 \centering
 \includegraphics[angle=270,width=0.8\linewidth]{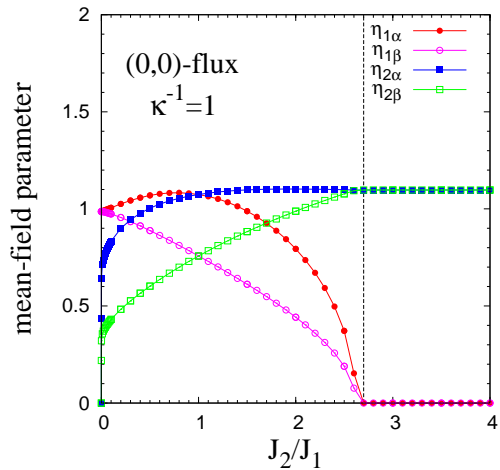}
 \caption{
 (Color online)
 Mean-field parameters of the long-range ordered states in (0,0)-flux ansatz at $\kappa^{-1}=1$.
 In the 120$^\circ$ N\'eel state at $J_2=0$,
 $\eta_{1\alpha}=\eta_{1\beta} \ne 0$ and $\eta_{2\alpha}=\eta_{2\beta} = 0$.
 The opposite is true for the collinear N\'eel state when $J_2/J_1 \geq 2.7$.
 The intermediate coplanar spiral state has a nonzero value for every mean-field parameter.
 \label{fig:MF_solutions_LRO}}
\end{figure}

The phase diagram for the (0,0)-flux ansatz is given in Fig. \ref{fig:0-flux-phase-diagram}.  In the quantum limit ($\kappa^{-1} \gg 1$), we have three spin liquid phases: the $2D$-$Z_2$, $3D$-$Z_2$, and $3D$-$U(1)$ states. In the classical limit ($\kappa^{-1} \ll 1$), there are three long-range orders: the 120$^\circ$ N\'eel, coplanar spiral, and collinear N\'eel states. The mean-field parameters of these states are depicted in Fig. \ref{fig:MF_solutions_SL} and \ref{fig:MF_solutions_LRO} for particular values of $\kappa$: $\kappa^{-1}=9$ for the spin liquid states and $\kappa^{-1}=1$ for the long-range ordered states. The six phases and the phase transitions among them are discussed below.

\paragraph{\underline{2D-$Z_2$ Spin Liquid}}
In the limit of small $J_2/J_1$, only the mean-field parameters in the triangular lattice plane, $\eta_{1\alpha}$ and $\eta_{1\beta}$, are non-zero, while the inter-plane parameters $\eta_{2\alpha} = \eta_{2\beta} = 0$. This yields a two-dimensional $Z_2$ spin liquid. Furthermore, since $\eta_{1\alpha} = \eta_{1\beta}$, it retains the three-fold rotational symmetry, and is in fact one of the bosonic spin liquids on the triangular lattice discovered in earlier studies.\cite{2006_wang} The minimum single-spinon gap occurs at the corners of the Brillouin zone, at the points $\pm$K. Furthermore, the $Z_2$ gauge flux excitations (visons) are also gapped.
\paragraph{\underline{3D-$Z_2$ Spin Liquid ({\it nematic} spin liquid)}} Upon
increasing $J_2/J_1$ further, the inter-plane couplings $\eta_{2\alpha}$ and
$\eta_{2\beta}$ become non-zero, along with the intra-plane ones,
$\eta_{1\alpha}$ and $\eta_{1\beta}$. Since these parameters all
have different magnitudes in general, they break the 120$^\circ$ lattice
rotation symmetry. We call such a spin liquid a {\em nematic} $Z_2$ spin
liquid. As the out-of-plane couplings are also non-zero, this is a gapped
three-dimensional $Z_2$ spin liquid. The minimum of the single-spinon gap
shifts, from the corners of the Brillouin zone at the $\pm$K points, to the
center of {an edge} of the Brillouin zone at the M
{point}, as $J_2/J_1$ increases.
\paragraph{\underline{3D-$U(1)$ Spin Liquid}}
When $J_2/J_1 \geq 1.3$ (for instance, at $\kappa^{-1} = 9$), a three-dimensional $U(1)$ spin liquid is stabilized. In this phase, the parameters in the triangular plane are zero, $\eta_{1\alpha} = \eta_{1\beta} = 0$, while the out-of-plane parameters, $\eta_{2\alpha} = \eta_{2\beta} \neq 0$. We immediately notice that the links on which the parameters are non-zero have a bipartite structure. Hence, we can define staggered $U(1)$ transformations of the following form:
\begin{align}
b_{i\mu}\to e^{i\phi}b_{i\mu}~~~~{\rm for}~ i\in {\rm A ~ sublattice},
\nonumber\\
b_{i\mu}\to e^{-i\phi}b_{i\mu}~~~~{\rm for}~ i\in {\rm B ~ sublattice},
\end{align}
for $\phi\in [0,2\pi)$. In other words, one can assign a positive gauge charge to one sublattice, and a negative gauge charge to the other.  This translates into the fact that the IGG is no longer $Z_2$, but $U(1)$, and  we have a three-dimensional $U(1)$ spin liquid. In three dimensions, such a phase may be stabilized, unlike the two-dimensional case. Analysis beyond mean-field shows that low-energy fluctuations of this $U(1)$ spin  liquid include two linearly dispersing photons, and hence this phase is gapless.\cite{2003_huse,2004_hermele,2005_bernier,2005_motrunich} The single-spinon gap reaches its minimum at the M {point}, the center of an edge of the Brillouin zone.

\paragraph{\underline{Long-range orders}}
Upon increasing $\kappa$, the spinon gap collapses, and condensation occurs,
leading to long-range magnetic order. Depending on the ratio of $J_2/J_1$,
different kinds of  magnetic orders are obtained. These are (1) {\em
120$^{\circ}$ non-collinear N\'eel order:} at $J_2 = 0$, we have decoupled
triangular lattices. For finite but moderate $\kappa$ ($\gtrsim 1/3$), this
supports 120$^{\circ}$ magnetic order, which is continuously connected to the
classical limit. (2) {\em coplanar spiral order:}  on increasing $J_2/J_1$, the
ordering wavevector for the spiral changes, and assumes an incommensurate value,
except at special values of $J_2/J_1$. However, the ordering is coplanar, and
the ordering wavevector is in the $x$-$y$ plane.\cite{Mulder_2010} While
the classical solution ($\kappa=\infty$) is degenerate in this regime, at finite
$\kappa$, a particular ordering wavevector ${\bf Q}$ is chosen through quantum
order by disorder, as shown in Fig. \ref{fig:Q_vector} {{(blue
line)}}. (3) {\em collinear N\'eel order:} At large $J_2/J_1$, the out-of-plane
couplings dominate, and 
since these couplings have a
bipartite structure, they stabilize a two-sublattice collinear N\'eel order.
Mean-field parameters for the above long-range ordered states at
$\kappa^{-1}=1$ are plotted in Fig. \ref{fig:MF_solutions_LRO}.

\subsubsection*{Phase Transitions.}

In Fig. \ref{fig:0-flux-phase-diagram}, the transition between the
incommensurate spiral and the $2D$-$Z_2$ spin liquid phases is first-order, while
the rest are second order. These second order transitions were studied by
Chubukov {\it et al.}\cite{1994_chubukov} in two dimension. Unlike the usual
Landau theory of phase transition, here the coarse grained action is not
written in terms of the magnetic order parameter, but in terms of the
low-energy spinons $b^{\dag}_{i\mu}$, the $U(1)$ gauge field, and the
gauge-charge-2 Higgs scalar field $\eta_{ij}$. Now, we will briefly describe
these transitions. Different phases depend on which of the above bosons are
condensed.

\begin{enumerate}
\item \underline{\em Neel - $3D$ $U(1)$} : The {\em collinear N\'eel phase} is described as a condensate of the spinons when the gauge-charge-2 Higgs field is gapped. The $U(1)$ gauge field is rendered massive through the Anderson-Higgs mechanism.\cite{1994_chubukov,2009_cenke} The transition from such a phase to a gapped $U(1)$ spin liquid is obtained by un-condensing the spinons while keeping the charge-2 Higgs field gapped. While this transition is continuous in the mean field level, it is known that fluctuation effects due to the gauge field may render this transition discontinuous in three dimensions.\cite{1974_halperin}

\item \underline{\em Neel - Spiral}: In contrast, the transition from the collinear N\'eel to spiral phase is obtained by condensing the charge-2 Higgs scalar in the background of the spinon condensate. 

\item \underline{\em $U(1)$ - $Z_2$}: Starting from the $U(1)$ spin liquid, one can condense the charge-2 Higgs scalar, and this breaks down the gauge group from $U(1)$ to $Z_2$. However, since the spinons are still gapped, this is a $Z_2$ spin liquid phase.

\item \underline{\em $Z_2$ - Spiral}: Finally, the transition from the $Z_2$ spin liquid to the spiral phase is obtained by condensing the spinons in the background of a charge-2 condensate, thereby completely gapping out the gauge fluctuations.
\end{enumerate}
\subsubsection{(0,$\pi$)-flux Ansatz\label{sec:pi-flux}}

\begin{figure}
 \centering
 \includegraphics[width=0.7\linewidth]{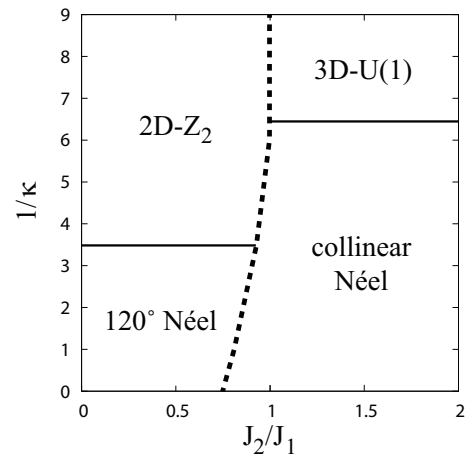}
 \caption{
 (Color online) Mean-field phase diagram of the (0,$\pi$)-flux ansatz.
 At small $J_2/J_1$, the 120$^{\circ}$ N\'eel (at small $\kappa^{-1}$) 
 and $2D$-$Z_2$ spin liquid (at large $\kappa^{-1}$) phases are found.
 Upon increasing $J_2/J_1$, 
 there is a first-order transition (denoted with dotted line) into 
 the collinear N\'eel (at small $\kappa^{-1}$) or $3D$-$U(1)$ spin liquid (at large $\kappa^{-1}$) states.
 The other transitions are second-order.
 In the phase diagram, 
 every phase preserves the lattice-rotation symmetry,
 unlike the (0,0)-flux phase diagram in Fig. \ref{fig:0-flux-phase-diagram}, 
 where there are intermediate phases with broken lattice-rotation symmetry,
 such as coplanar spiral and $3D$-$Z_2$ spin liquid states.
 \label{fig:pi-flux-phase-diagram}}
\end{figure}

The phase diagram for the (0,$\pi$)-flux phase is shown in Fig. \ref{fig:pi-flux-phase-diagram}. No new phases are present; however, the incommensurate spiral phase at large $\kappa$  and $3D$-$Z_2$ spin liquid at small $\kappa$ are missing in the intermediate $J_2/J_1$ range.

We note that none of the phases in the phase diagram (Fig. \ref{fig:pi-flux-phase-diagram}) break the lattice rotational symmetry. In addition, the mean-field parameters in the triangular planes are never simultaneously nonzero along with the inter-plane parameters. At small $J_2/J_1$, the 120$^{\circ}$ N\'eel (at large $\kappa$) and $2D$-$Z_2$ spin liquid (at small $\kappa$) phases are found. Upon increasing $J_2/J_1$, there is a first-order transition into the collinear N\'eel (at large $\kappa$) or $3D$-$U(1)$ spin liquid (at small $\kappa$) states. This is unlike the (0,0)-flux phase diagram, where there are intermediate incommensurate spiral and $3D$-$Z_2$ spin liquid phases. Hence, in the $\kappa\rightarrow\infty$ limit, the magnetically ordered phase is not the same for intermediate $J_2/J_1$ as obtained from the classical analysis.\cite{Mulder_2010}

\subsubsection*{(0,0)-flux ansatz vs. (0,$\pi$)-flux ansatz}

\begin{figure}
 \centering
 \includegraphics[angle=270,width=0.45\linewidth]{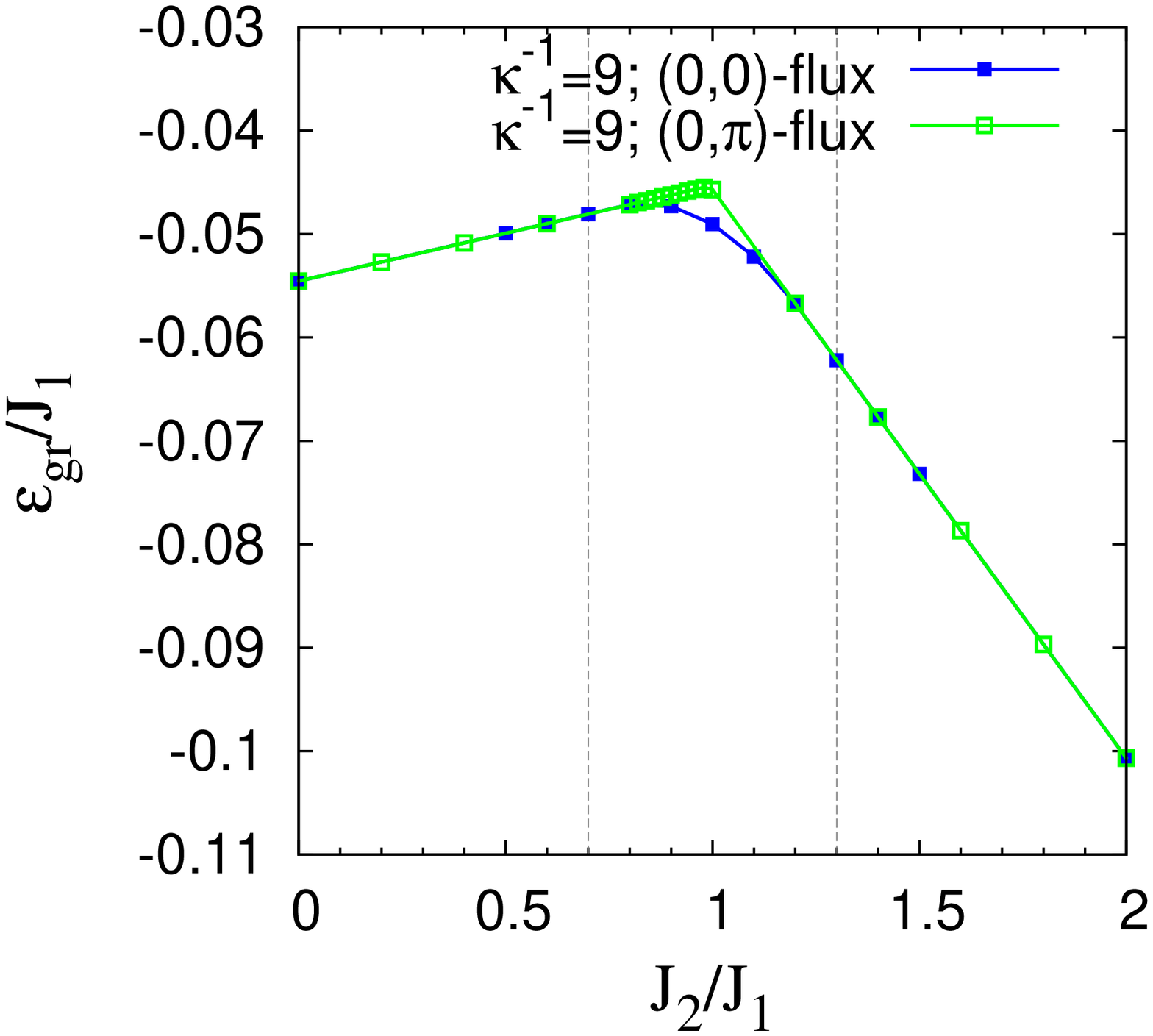}
 \includegraphics[angle=270,width=0.45\linewidth]{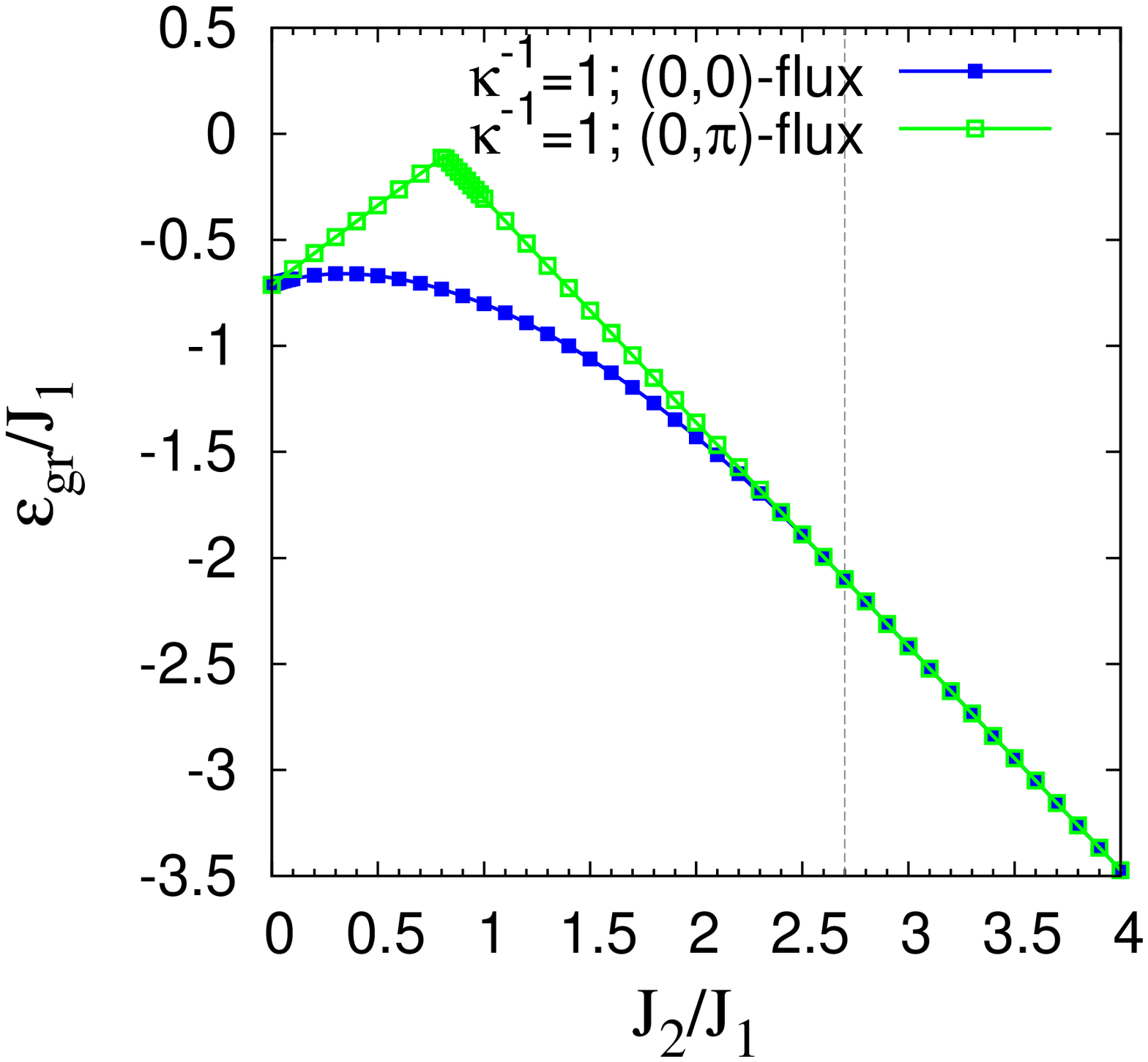}
 \caption{
 (Color online)
 Ground state energy per site ($\epsilon_{gr}$) of the (0,0)-flux and (0,$\pi$)-flux ans\"{a}tze.
 $\epsilon_{gr}$ is defined in (\ref{eq:e_gr of H_MF}).
 Left: $\kappa^{-1}=9$ (spin liquid regime).
 Right: $\kappa^{-1}=1$ (long-range order regime).
 The {above} plots of $\epsilon_{gr}$ for both ans\"atze {show} that the (0,0)-flux ansatz {{(blue)}} {{is generally more stable than}} the (0,$\pi$)-flux ansatz {{(green)}}.
 \label{fig:e_gr}}
\end{figure}

Before moving to next subsection, we compare the (0,0)-flux and (0,$\pi$)-flux
ans\"atze in terms of the ground state energy and symmetry. Figure
\ref{fig:e_gr} shows their ground state energy per site as a function of
$J_2/J_1$ for two particular cases: $\kappa^{-1}=9$ and $\kappa^{-1}=1$. The
ground state energy per site ($\epsilon_{gr}$) is defined in (\ref{eq:e_gr of
H_MF}). From both cases in the figure, it is clearly seen that the (0,0)-flux
ansatz (blue)  is generally more stable than the (0,$\pi$)-flux ansatz (green). In
fact, there are two parameter regions where the two ans\"atze have the same
energy: the small $J_2/J_1$ region and large $J_2/J_1$ region. In these
regions, they have physically identical mean-field solutions, which preserve the
lattice-rotation symmetry, in the $2D$-$Z_2$, $3D$-$U(1)$, 120$^\circ$ N\'eel,
and collinear N\'eel states. However, when $J_1$ and $J_2$ are comparable, two
ans\"atze have different mean-field solutions with different symmetries  as
discussed in Sec. \ref{sec:0-flux} and \ref{sec:pi-flux}. In the (0,0)-flux ansatz,
the $3D$-$Z_2$ and coplanar spiral states with
broken rotation symmetry 
are stabilized. However, the (0,$\pi$)-flux ansatz allows only the other
states with the rotation symmetry. Energetically, the broken symmetry states
are more stable than the symmetry-preserving states, as shown in Fig.
\ref{fig:e_gr}.

From {the} above energy comparisons, we find that (i) the (0,0)-flux
ansatz provides more stable states compared to the (0,$\pi$)-flux ansatz and (ii)
the system is stabilized by breaking the lattice-rotation symmetry when
the system is frustrated due to competing interactions $J_1$ and $J_2$.
\subsection{Phase Diagram for SG$_2$}

The phase diagram obtained for the SG$_2$ ansatz is the same as for the (0,$\pi$)-flux ansatz for SG$_1$, since none of the phases obtained break rotational symmetry.  Consequently, both of the ans\"atze have identical mean-field solutions.

\section{Two-spinon excitation spectra\label{sec:two-spinon spectra}}

\begin{figure*}[ht]
 \centering
 \includegraphics[angle=270,width=0.18\linewidth]{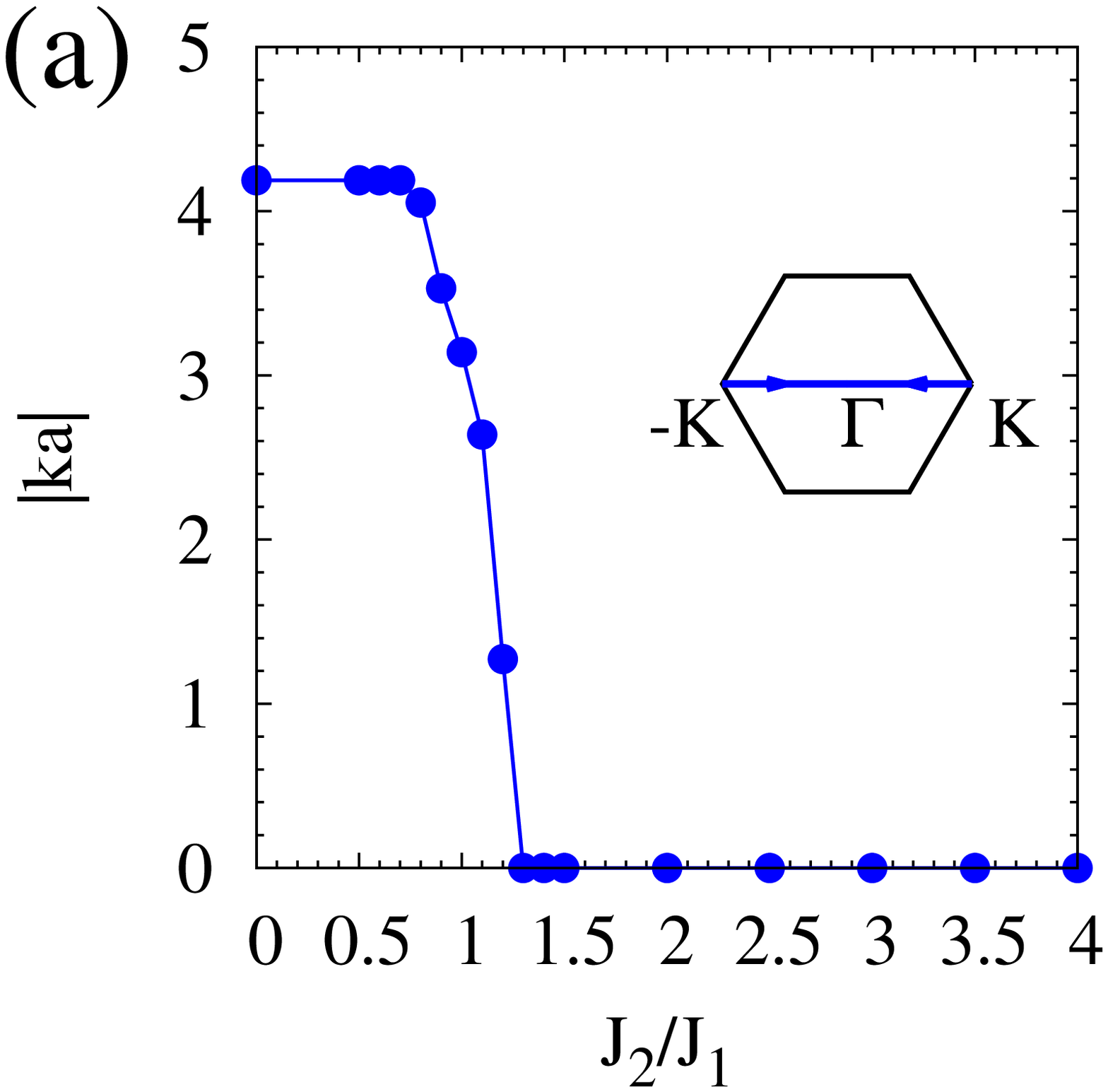}
 \includegraphics[bb=80 170 554 670,clip,width=0.18\linewidth,angle=270]{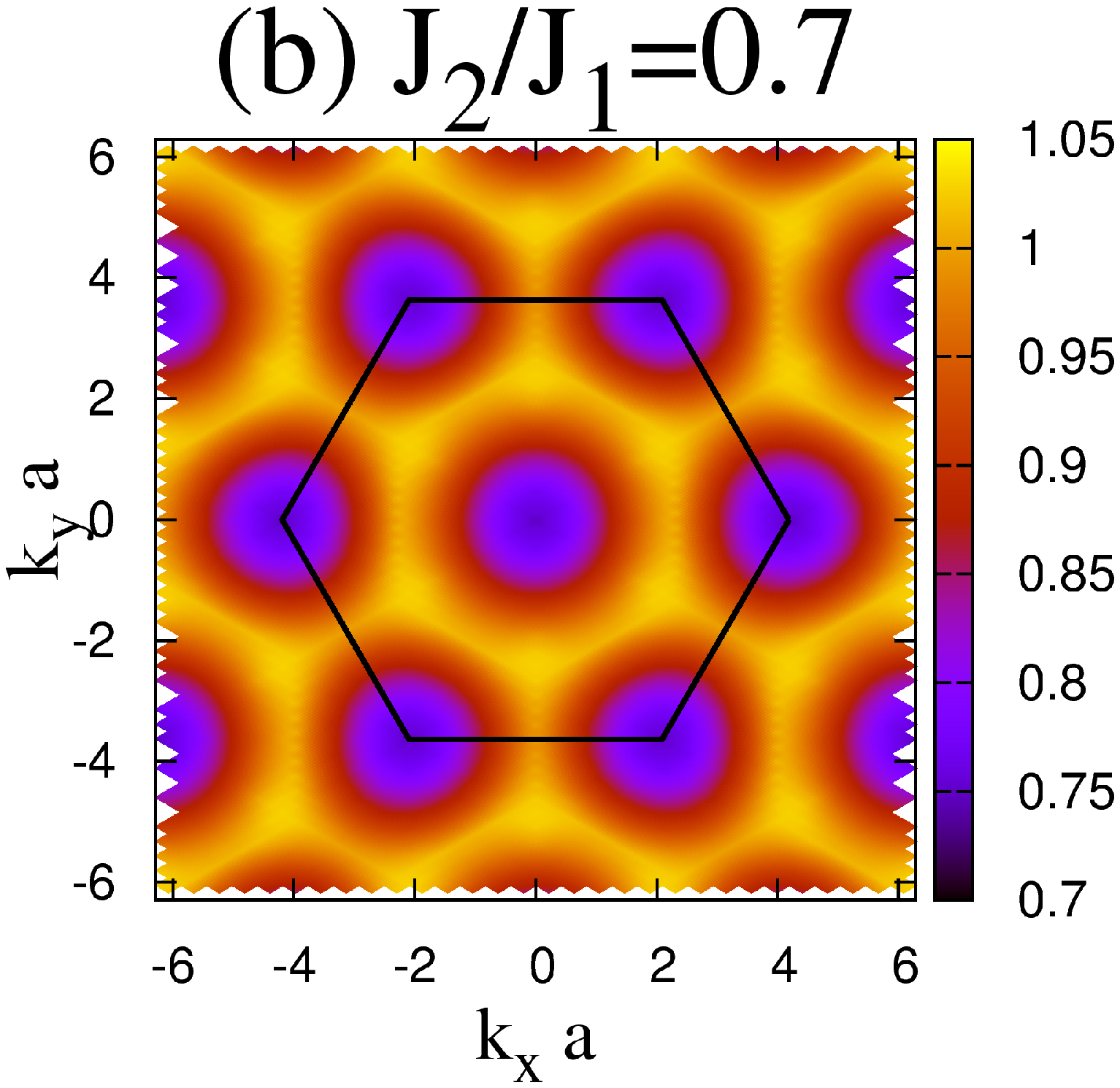}
 \includegraphics[bb=80 170 554 670,clip,width=0.18\linewidth,angle=270]{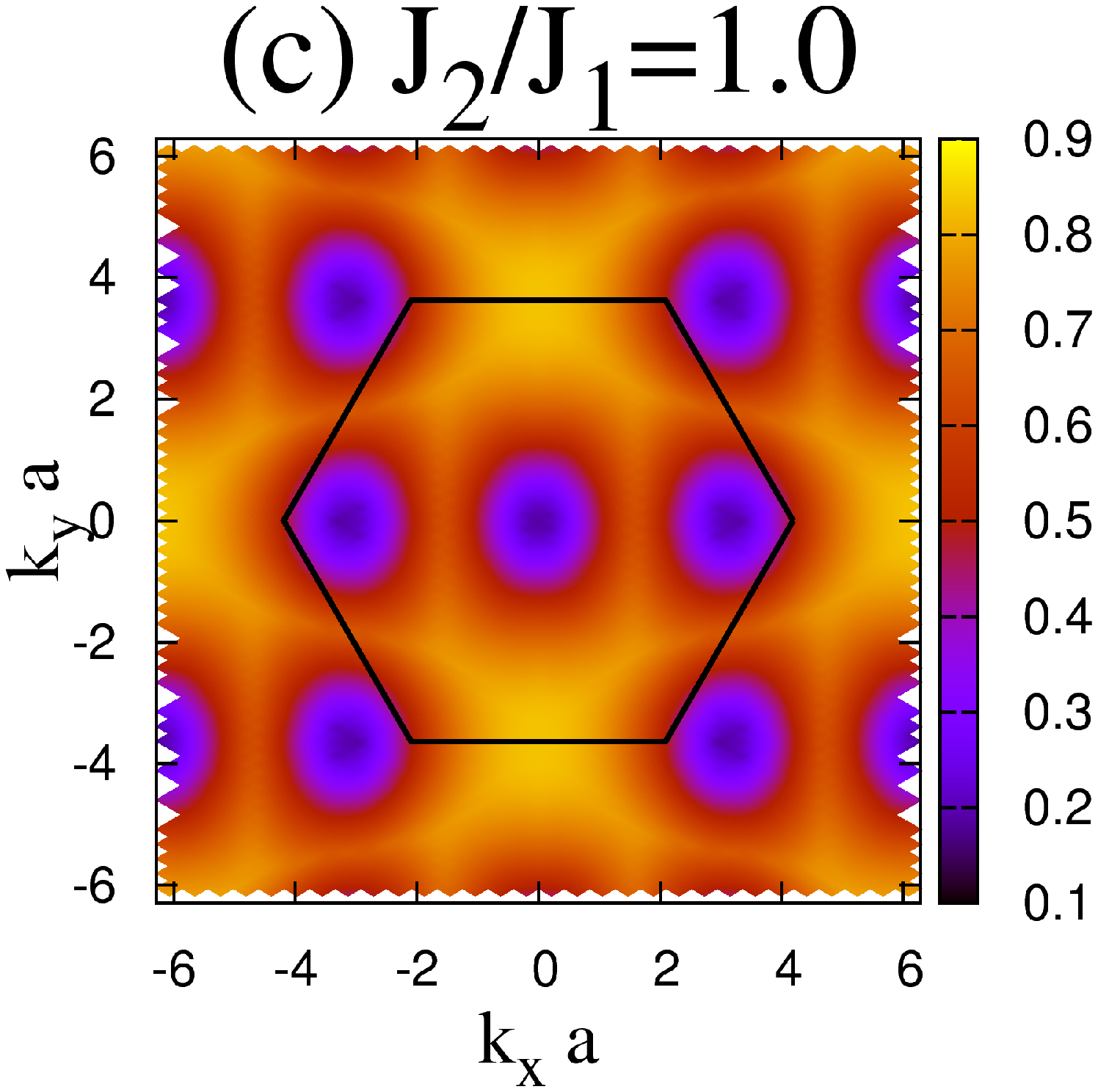}
 \includegraphics[bb=80 170 554 670,clip,width=0.18\linewidth,angle=270]{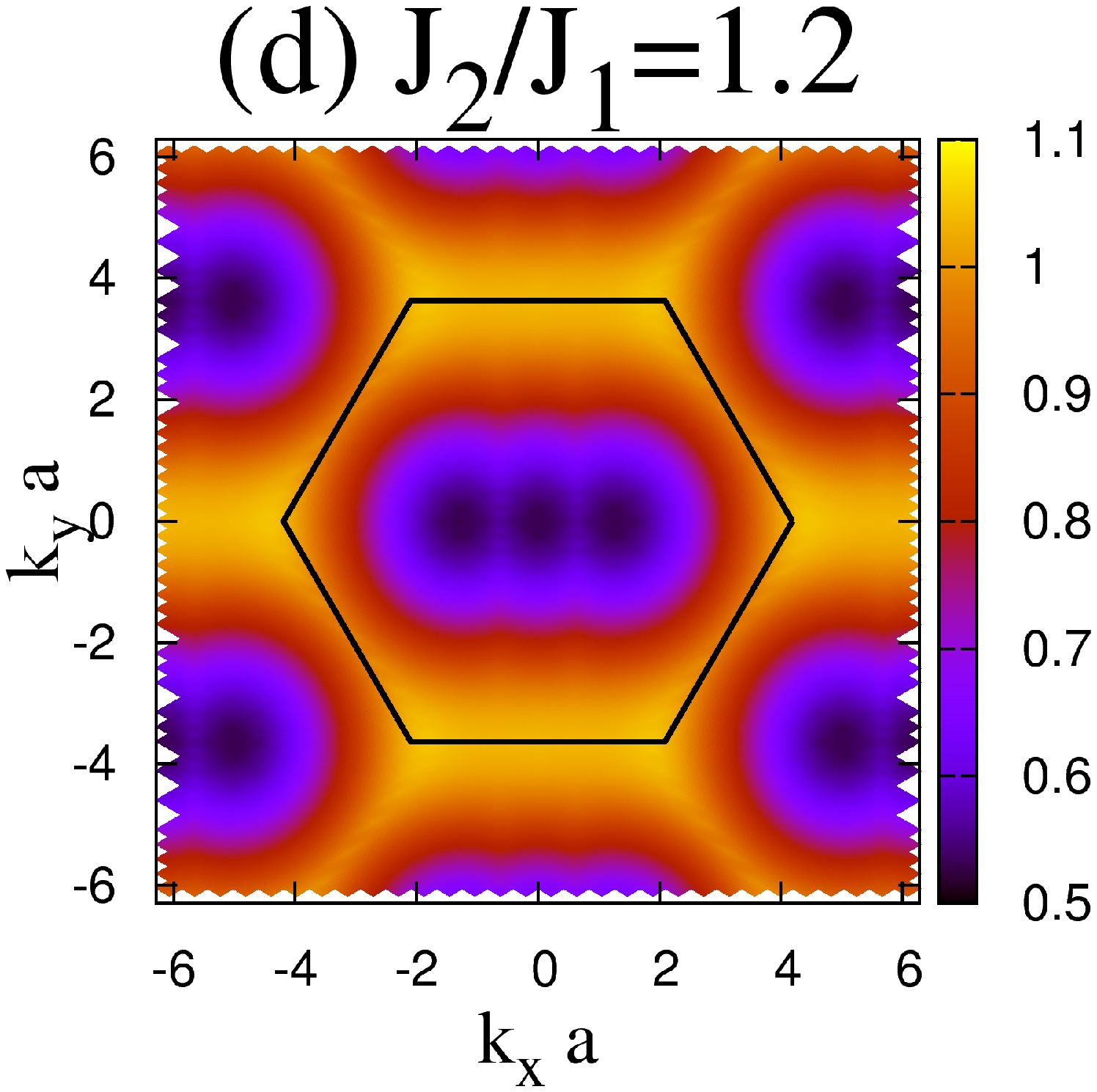}
 \includegraphics[bb=80 170 554 670,clip,width=0.18\linewidth,angle=270]{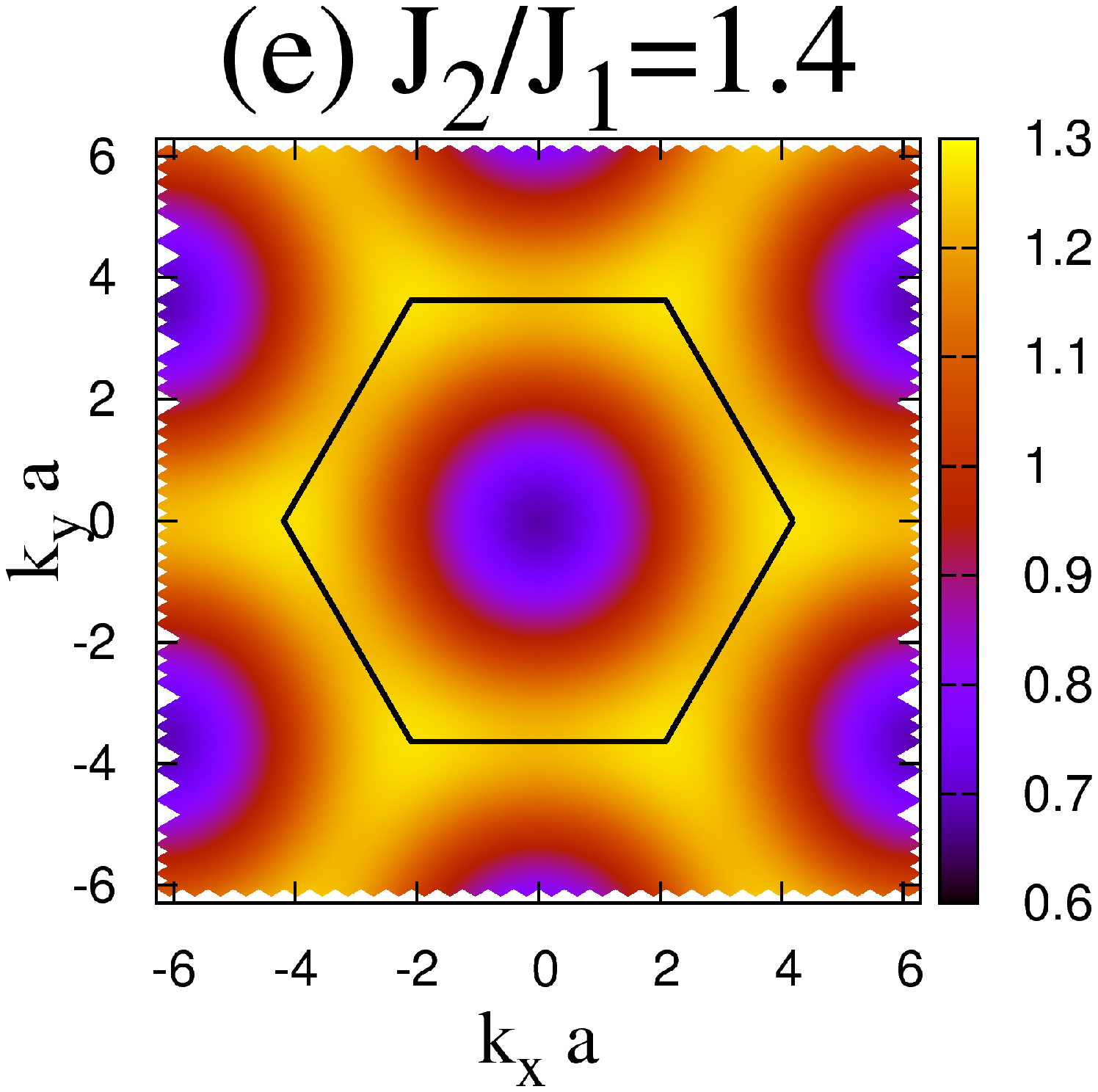}
 \caption{
(Color online)
 The lower edge of the two-spinon excitations, $E_2({\bf k})/J_1$, of the (0,0)-flux spin liquid states at $\kappa^{-1}=9$.
 $E_2({\bf k})$ is defined in (\ref{eq:two-spinon excitation}).
 We plot the plane $k_z=0$, which contains the minima of these excitations.
 (a) The positions of the minima in $E_2({\bf k})$ in the first Brillouin zone as a function of $J_2/J_1$.
 (b) The $2D$-$Z_2$ spin liquid.
 (c) $\sim$ (d) The $3D$-$Z_2$ spin liquids.
 (e) The $3D$-$U(1)$ spin liquid.
 The two minimum points at $\pm$K gradually move toward the $\Gamma$ point and then merge at the point 
 as $J_2/J_1$ increases.
 In each plot, the hexagon denotes the first Brillouin zone at $k_z=0$.
 \label{fig:two spinon dispersion}}
\end{figure*}

Having already discussed the ground state properties of the various spin liquid
phases in the previous section, we briefly discuss the two-spinon
excitation spectra for various spin liquid states that we find for the (0,0)-flux
ansatz with SG$_1$. Such a two-spinon excitation spectrum is gauge
invariant (unlike the spingle spinon spectrum) and is proportional to the
spin-structure factor measured in the neutron scattering
experiments.\cite{2008_lawler, 2001_chung} Hence,
it reflects
the physical symmetries broken
in a phase.
In our case,
the two-spinon spectrum provides a valuable information for investigating the
symmetries of a spin liquid state, especially the broken lattice-rotation
symmetry in the nematic spin liquid state.

The lower edge of the two-spinon spectrum is given by\cite{2008_lawler,2001_chung}
\begin{align}
 E_{2} ({\bf k}) = {\underset{\bf p}{\textup{min}}}{ \left[ \epsilon ({\bf p}) + \epsilon ({\bf k}-{\bf p}) \right] },
 \label{eq:two-spinon excitation}
\end{align}
where $\epsilon ({\bf k})=\textup{min} \{ \omega_{1}({\bf k}), \cdots , \omega_{n_s}({\bf
k}) \}$  and $\omega_{n}({\bf k})$ $(n=1,\cdots,n_s)$ are energy bands of the single-spinon
excitations. For more details about $\omega_{n}({\bf k})$, the reader is referred
to Appendix \ref{appendix:mean-field Hamiltonians}. $E_{2} ({\bf k})$ of the
(0,0)-flux state is plotted in Fig. \ref{fig:two spinon dispersion} for several
values of
$J_2/J_1$ at $\kappa^{-1}=9$. The figure clearly shows that the $3D$-$Z_2$
state breaks the 120$^\circ$ lattice-rotation symmetry, (see (c),(d)) whereas
the $2D$-$Z_2$ and $3D$-$U(1)$ preserve the rotation symmetry (see (b),(e)). In
addition, it is observed that $E_{2} ({\bf k})$ continuously changes as
$J_2/J_1$ increases. The $2D$-$Z_2$ spin liquid state $(J_2/J_1 \leq 0.7)$ has
the minima of $E_{2} ({\bf k})$ at the $\pm$K points. In the $3D$-$Z_2$ spin
liquid states, $(0.7 < J_2/J_1 < 1.3)$ the two minima in the two-spinon
spectrum gradually move toward the $\Gamma$ point as $J_2/J_1$ increases. When
$J_2/J_1=1.3$, the two minima merge at the $\Gamma$ point, which is the minimum
of the two-spinon spectrum of the $3D$-$U(1)$ spin liquid $(J_2/J_1 \geq 1.3)$.
The continuous change in the minimum points is depicted in Fig. \ref{fig:two
spinon dispersion} (a). The minimum points of each spin liquid state are
consistent with the ordering wave vectors of the corresponding long range order
state.

The existence of the (0,0)-flux $3D$-$Z_2$ spin liquid state is an example of a {\em nematic spin liquid} that breaks the $120^\circ$ rotational symmetry, but retains time-reversal, spin-rotational, and all other symmetries of the lattice. However, since the spinons may remain in a deconfined phase, such a state retains the fractionalized spin-$1/2$ excitations and topological order that comprise the exotic {behaviors} of symmetric spin liquids. Its energetic favorability within mean-field theory, combined with its necessity to describe the spiral order, {indicates} that such a state may be relevant for a model
with the 6H-B structure. Although the Schwinger boson mean-field theory predicts that spiral order is stabilized over a large range of $\kappa$, such a nematic spin liquid serves as a starting point for understanding spin-disordered ground states of the frustrated 6H-B structure.

\section{Summary and outlook\label{sec:summary-outlook}}

In this work, we have analyzed the possible three dimensional bosonic spin
liquids on an alternately stacked triangular lattice structure. 
For the frustrated Heisenberg
Hamiltonian with nearest neighbor intra- and inter-plane
couplings, we find several three dimensional quantum spin-liquid
phases for small values of $\kappa~(\sim 2S)$. 
A central finding of our analysis is a 3D {\it nematic} $Z_2$ spin liquid that breaks
lattice rotation symmetry that is reflected in the spin-structure factor.
Interestingly, our Schwinger boson mean field theory calculation shows that
this spin liquid is not only energetically more stable (within self-consistent
mean-field theory) than an isotropic one in the same parameter regime, but is
also naturally connected to the classical limit 
of
coplanar spiral
order. We note that spiral order also breaks lattice rotation symmetries along
with spin rotation. It is interesting that this spontaneously broken lattice
rotation persists in the nematic spin liquid, even though the spin-rotation
invariance is restored across the transition. Although our PSG analysis is aimed
to capture $Z_2$ spin liquids, we identify a bipartite structure  of the
mean-field parameters in one of the spin liquid phases. The resultant phase
actually turns out to be a gapped U(1) spin liquid in three dimensions. 

In the present self-consistent mean-field calculations of the given spin-model,
the maximum $\kappa\sim 2S$ for which the various spin liquids are stable is
rather small compared to the physical value of the spin. However, the
spin liquids found here may be stabilized by further neighbor interactions or
multi-spin exchange processes for higher values of $\kappa$. It
would be interesting to see the fate of the nematic spin liquid in such models.
Another interesting feature of the phase diagram {is}
the existence of
various unconventional quantum phase transitions between the magnetically
ordered and disordered phases. Most of these transitions turn out to be
continuous within our mean-field theory. We identify these transitions by
extending the existing framework\cite{1994_chubukov,2009_cenke} in terms of
condensation of spinons to three spatial dimensions. 

It would be interesting to explore,
in addition to Ba$_3$NiSb$_2$O$_9$,
other possible Mott-insulators on three dimensional AB-stacked triangular
lattices that may stabilize one or more
of the
spin liquid {phases} described here. Of particular interest would be
the materials close to metal-insulator transition where charge fluctuations may
enhance further neighbor interactions and multi-spin processes.

\begin{acknowledgments}
We thank R. Schaffer for useful discussions. 
YBK acknowledges the support and hospitality from the Aspen Center for Physics, funded by NSF grant PHY-1066293, and the KITP, funded by NSF grant PHY–1125915.
This research was supported by the NSERC, CIFAR, and Centre for Quantum Materials at the University of Toronto.
{{The numerical computations were done in SciNet at the University of Toronto.}}
\end{acknowledgments}

\appendix

\section{Projective Symmetry Group Construction I{\label{appendix:PSG-construction}}}

We split the determination of the PSG into two parts; the first described the procedure and results, leaving most
of the details to the second.

In this Appendix,
we introduce twenty five multiplication rules among the seven symmetry operations given in (\ref{eq:space-group-transformations}),
construct the PSG corresponding to the three translations, and
present the final PSG for both symmetry groups SG$_1$ and SG$_2$.
We leave the derivation of the PSG elements for the other 
symmetry group operations to Appendix \ref{appendix:PSG-construction-2}.

The multiplication rules among the seven symmetry operations in (\ref{eq:space-group-transformations}) are given by the following:
\begin{subequations}
\label{eq:constraints_T_n}
\begin{eqnarray}
 &&
 T_1 T_2 = T_2 T_1,
 \\
 &&
 T_2 T_3 = T_3 T_2,
 \\
 &&
 T_1 T_3 = T_3 T_1,
\end{eqnarray}
\end{subequations}
\begin{subequations}
\label{eq:constraints_Pi_1}
\begin{eqnarray}
 &&
 T_1 \Pi_{1} = \Pi_{1} T_1^{-1},
 \\
 &&
 T_2 \Pi_{1} = \Pi_{1} T_1 T_2,
 \\
 &&
 T_3 \Pi_{1} = \Pi_{1} T_3,
 \\
 &&
 \Pi_{1}^{2} = I,
\end{eqnarray}
\end{subequations}
\begin{subequations}
\label{eq:constraints_Pi_2}
\begin{eqnarray}
 &&
 T_1 \Pi_{2} = \Pi_{2} T_1,
 \\
 &&
 T_2 \Pi_{2} = \Pi_{2} T_2,
 \\
 &&
 T_3 \Pi_{2} = \Pi_{2} T_3^{-1},
 \\
 &&
 \Pi_{2}^{2} = I,
 \\
 &&
 \Pi_{1} \Pi_{2} = \Pi_{2} \Pi_{1},
\end{eqnarray}
\end{subequations}
\begin{subequations}
\label{eq:constraints_Xi}
\begin{eqnarray}
 &&
 T_1 \Xi = \Xi T_1^{-1},
 \\
 &&
 T_2 \Xi = \Xi T_2^{-1},
 \\
 &&
 T_3 \Xi = \Xi T_3^{-1},
 \\
 &&
 \Xi^{2} = I,
 \\
 &&
 \Pi_{1} \Xi = \Xi \Pi_{1},
 \\
 &&
 \Pi_{2} \Xi = \Xi T_3 \Pi_{2}.
\end{eqnarray}
\end{subequations}
\begin{subequations}
\label{eq:constraints_R}
\begin{eqnarray}
 &&
 T_1 R = R T_1^{-1} T_2^{-1},
 \\
 &&
 T_2 R = R T_1,
 \\
 &&
 T_3 R = R T_3,
 \\
 &&
 R^3 = I,
 \\
 &&
 R \Pi_{1} = \Pi_{1} R^{-1},
 \\
 &&
 R \Pi_{2} = \Pi_{2} R,
 \\
 &&
 R \Xi = \Xi T_1^{-1} T_2^{-1} R,
\end{eqnarray}
\end{subequations}
where $I$ is the identity operator.
The above multiplication rules provide constraints on the PSG of the symmetry operations.

We now construct the PSG for the subgroup of translations defined by
the equation (\ref{eq:constraints_T_n}), re-writing it as
\begin{subequations}
\begin{eqnarray}
 &&
 T_1^{-1} T_2 T_1 T_2^{-1} = I,
 \label{eq:SG_T_constraint_1}
 \\
 &&
 T_2^{-1} T_3 T_2 T_3^{-1} = I,
 \label{eq:SG_T_constraint_2}
 \\
 &&
 T_1^{-1} T_3 T_1 T_3^{-1} = I.
 \label{eq:SG_T_constraint_3}
\end{eqnarray}
\end{subequations}
By considering the equivalent expression for PSG elements, as in
(\ref{eq:identity-to-igg}), from (\ref{eq:SG_T_constraint_1})
we obtain
\begin{eqnarray}
 &&
 (G_{T_1} T_1)^{-1}
 (G_{T_2} T_2)
 (G_{T_1} T_1)
 (G_{T_2} T_2)^{-1}
 \nonumber\\
 & = &
 (T_1^{-1} G_{T_1}^{-1} T_1)
 \cdot
 (T_1^{-1} G_{T_2} T_1)
 \nonumber\\
 &&
 \cdot
 ((T_1 T_2^{-1})^{-1} G_{T_1} (T_1 T_2^{-1}))
 \cdot
 G_{T_2}^{-1}
 \nonumber\\
 & \in & 
 \mathrm{IGG}=\textup{Z}_2,
 \nonumber
\end{eqnarray}
where $G_X$ ($X=T_1,T_2$) is the gauge transformation 
associated with the operation $X$ in the SG.
Under 
$G_X$, the boson operator acquires the phase $e^{i \phi_X ({\bf r})}$.
The phase of the above equation can be written as
\begin{eqnarray}
 &&
 -\phi_{T_1}[T_1({\bf r})]
 +\phi_{T_2}[T_1({\bf r})]
 \nonumber\\
 &&
 +\phi_{T_1}[T_1 T_2^{-1}({\bf r})]
 -\phi_{T_2}[{\bf r}]
 = n_1 \pi,
 \label{eq:PSG_T_constraint_1}
\end{eqnarray}
where $n_1 \in \{ 0,1 \}$ and $n_1=0$ ($n_1=1$) corresponds to the element $+1$ ($-1$) in the Z$_2$ gauge group.
Equation (\ref{eq:PSG_T_constraint_1}) is the first constraint on the PSG.
Whenever we derive a constraint on the PSG from a multiplication rule among the symmetry operations,
we employ an integer variable, like $n_1$ in the above equation.
Hereafter, $n_i ~ (i=1,\cdots,25)$ is regarded as an integer variable, which is either 0 or 1.
It must be noted that in all phase equations like (\ref{eq:PSG_T_constraint_1}), 
equality is to be understood as modulo $2\pi$, due to the 2$\pi$ periodicity in the phases.

Determining the possible solutions of the constraint
equations like (\ref{eq:PSG_T_constraint_1}) can be
facilitated by an appropriate choice of gauge fixing.
For the first gauge fixing, we set
\begin{align}
 \phi_{T_1}(r_1,r_2,r_3)_p=0,
 \label{eq:PSG_T_1}
\end{align}
for both sublattices ($p=A,B$).
Then, (\ref{eq:PSG_T_constraint_1}) is reduced to 
\begin{align}
 \phi_{T_2}(r_1,r_2,r_3)_p = \phi_{T_2}(0,r_2,r_3)_p + n_1 \pi r_1.
\end{align}
Taking another gauge fixing
\begin{align}
 \phi_{T_2}(0,r_2,r_3)_p = 0,
 \label{eq:gauge_fixing_2}
\end{align}
we obtain
\begin{align}
 \phi_{T_2}(r_1,r_2,r_3)_p = n_1 \pi r_1.
 \label{eq:PSG_T_2}
\end{align}
In a similar way, $\phi_{T_3}({\bf r})$ is determined from (\ref{eq:SG_T_constraint_2}) and (\ref{eq:SG_T_constraint_3}).
If we use (\ref{eq:PSG_T_1}) and (\ref{eq:PSG_T_2}) and take the gauge-fixing
\begin{align}
 \phi_{T_3}(0,0,r_3)_p = 0,
 \label{eq:gauge_fixing_3}
\end{align}
then we are lead to
\begin{align}
{
 \phi_{T_3}(r_1,r_2,r_3)_p = n_2 \pi r_2 + n_3 \pi r_1.
  \label{eq:PSG_T_3}
}
\end{align}

We must ensure to fix the gauge freedom so as
to satisfy (\ref{eq:PSG_T_1}), (\ref{eq:gauge_fixing_2}),
(\ref{eq:gauge_fixing_3}) at the same time.
The gauge-fixings can be realized in the following way.\cite{Choy_PSG_2009}
Suppose that $\phi_{T_n}^{(0)}~(n=1,2,3)$ is the initial choice for the 
phase of $G_{T_n}$.
Then, we consider a series of gauge transformations $G_m ~ (m=1,2,3)$,
so that
\begin{subequations}
\begin{align}
 \phi_{T_n}^{(0)} \xrightarrow{G_1} \phi_{T_n}^{(1)} \xrightarrow{G_2} \phi_{T_n}^{(2)} \xrightarrow{G_3} \phi_{T_n}^{(3)},
\end{align}
where $G_m=e^{i\phi_{G_m}}~(m=1,2,3)$ is defined as follows:
\begin{align}
 \phi_{G_m}(r_1,r_2,r_3)_p
 =
 &-& \sum_{i=-\infty}^{r_1} \phi_{T_1}^{(m-1)}(i,r_2,r_3)_p
 \nonumber\\
 &-& \sum_{j=-\infty}^{r_2} \phi_{T_2}^{(m-1)}(0,j,r_3)_p 
 \nonumber\\
 &-& \sum_{k=-\infty}^{r_3} \phi_{T_3}^{(m-1)}(0,0,k)_p.
\end{align}
 \label{eq:gauge_fixing_123_realization}
\end{subequations}
It is important to note that $\phi_{G_m}$ depends on $\phi_{T_n}^{(m-1)}$.
To determine $\phi_{T_n}^{(m)}~(m=1,2,3)$, it is necessary to understand how an
arbitrary gauge transformation changes $G_X$.
It can be understood
from the fact that our mean-field ansatz is invariant under the PSG, $\{G_X
X\}$.
Then, if we take a gauge transformation $G$, the transformed mean-field ansatz is invariant under $G \cdot G_X X \cdot G^{-1} = \tilde{G}_X \cdot X$, where $\tilde{G}_X = G \cdot G_X \cdot X G^{-1} X^{-1}$.
This gives us a new PSG, $\{ \tilde{G}_X \cdot X \}$.
Then, under the transformation $G$,
$\phi_X [{\bf r}]$ changes in following way.
\begin{align}
 \phi_X [{\bf r}]
 \xrightarrow{G}
 \phi_G [{\bf r}] + \phi_X [{\bf r}] - \phi_G [X^{-1}({\bf r})].
\end{align}
According to above transformation rule, 
$\phi_X [{\bf r}]$ acquires the additional phase $\phi_G [{\bf r}] - \phi_G [X^{-1}({\bf r})]$ under the transformation $G$.
Applying the above rule to each transformation in (\ref{eq:gauge_fixing_123_realization}),
we find that the gauge fixings (\ref{eq:PSG_T_1}), (\ref{eq:gauge_fixing_2}), (\ref{eq:gauge_fixing_3}) are all satisfied.
As seen in the definition of (\ref{eq:gauge_fixing_123_realization}), the above gauge-fixing is
defined for open boundary conditions.
Periodic boundary conditions 
introduce additional subtleties, so we consider the PSG for
systems with open boundary conditions, with the understanding that the particular boundary condition
will be irrelevant in the thermodynamic limit.

Readers who are interested in detailed derivation of $G_X$ for the rest of the
symmetry group are referred to Appendix \ref{appendix:PSG-construction-2}.
We present the phases $\phi_X$ of $G_X$ for the full PSG below.

\underline{PSG for SG$_1$}:

\begin{subequations}
\label{eq:PSG_for_SG_1}
\begin{eqnarray}
 \phi_{T_{1,3}} (r_1,r_2,r_3)_p 
 &=&
 0,
 \label{eq:PSG_for_SG_1_T_1}
 \\
 \phi_{T_2} (r_1,r_2,r_3)_p 
 &=&
 n_1 \pi r_1,
 \label{eq:PSG_for_SG_1_T_23}
 \\
 \phi_{\Pi_1} (r_1,r_2,r_3)_p
 &=&
 \left( \frac{1}{2} + \delta_{p,B} \right) \pi - \frac{1}{2} n_1 \pi r_2 (r_2 - 1),
 \label{eq:PSG_for_SG_1_Pi_1}
 \nonumber\\
 \\
 \phi_{\Pi_2} (r_1,r_2,r_3)_p
 &=&
 0,
 \label{eq:PSG_for_SG_1_Pi_2}
 \\
 \phi_{\Xi} (r_1,r_2,r_3)_p 
 &=& 
 \frac{\pi}{2} + n_1 \pi r_1,
 \label{eq:PSG_for_SG_1_Xi}
\end{eqnarray}
\end{subequations}
where $n_1=0,1$.

Considering the remaining rotation $R$,
we obtain the PSG for the SG$_2$.
The rotational symmetry leads to $n_1=0$ in (\ref{eq:PSG_for_SG_1}) 
as shown in Appendix \ref{appendix:PSG-construction-2-rotation}.

\underline{PSG for SG$_2$}:

\begin{subequations}
\label{eq:PSG_for_SG_2}
\begin{eqnarray}
 \phi_{T_{1,2,3}} (r_1,r_2,r_3)_p 
 &=&
 0,
 \\
 \phi_{\Pi_1} (r_1,r_2,r_3)_p
 &=&
 \left( \frac{1}{2} + \delta_{p,B} \right) \pi,
 \\
 \phi_{\Pi_2} (r_1,r_2,r_3)_p
 &=&
 0,
 \\
 \phi_{\Xi} (r_1,r_2,r_3)_p 
 &=& 
 \frac{\pi}{2},
 \\
 \phi_{R} (r_1,r_2,r_3)_p 
 &=& 0.
\end{eqnarray}
\end{subequations}

\section{Projective Symmetry Group Construction II{\label{appendix:PSG-construction-2}}}

In this section, we derive the transformation $G_X = e^{i\phi_X}$ for the
remaining elements of SG$_1$ and SG$_2$ beyond the translations.

\subsection{$\phi_{\Pi_1} (r_1,r_2,r_3)_p$}

Rewriting (\ref{eq:constraints_Pi_1}) into constraints like (\ref{eq:PSG_T_constraint_1}),
and solving the resultant equations, 
we obtain
\begin{eqnarray}
 &&
 \phi_{\Pi_1} (r_1,r_2,r_3)_p
 \nonumber\\
 &=&
 \phi_{\Pi_1} (0,0,0)_p
 +
 n_4 \pi r_1
 +
 n_5 \pi r_2
 +
 n_6 \pi r_3
 \nonumber\\
 &-&
 \frac{n_1}{2} \pi r_2 ( r_2 - 1)
 -
 n_3 \pi r_2 r_3,
 \label{eq:PSG_Pi_1_eq_1}
\end{eqnarray}
and
\begin{subequations}
\label{eq:PSG_Pi_1_eq_2}
\begin{eqnarray}
 &&
 2 \phi_{\Pi_1} (0,0,0)_p = n_7 \pi,
 \\
 &&
 n_4 = 0.
\end{eqnarray}
\end{subequations}
Of the above equations,
(\ref{eq:PSG_Pi_1_eq_1}) is determined by (\ref{eq:constraints_Pi_1}a,b,c)
and (\ref{eq:PSG_Pi_1_eq_2}) by (\ref{eq:constraints_Pi_1}d).
As mentioned earlier, integer variables ($n_i$) are introduced 
when multiplication rules among the symmetry operations are used to place a constraint on the PSG.
In the above expressions,
the integer variable $n_5$ can be eliminated by the gauge transformation $G_4=e^{i\phi_{G_4}}$,
where
\begin{align}
 \phi_{G_4}(r_1,r_2,r_3)_p = \pi r_1.
 \nonumber
\end{align}
Under this transformation,
\begin{eqnarray}
 &&
 \phi_{T_n}(r_1,r_2,r_3)_p \rightarrow \phi_{T_n}(r_1,r_2,r_3)_p + \delta_{n,1} \pi ~ (n=1,2,3),
 \nonumber\\
 &&
 \phi_{\Pi_1}(r_1,r_2,r_3)_p \rightarrow \phi_{\Pi_1}(r_1,r_2,r_3)_p - \pi r_2.
 \nonumber
\end{eqnarray}
In fact, the additional constant $ \delta_{n,1} \pi $ in $\phi_{T_n}$ can be ignored
since we may add a site-independent constant $\pi$ to $\phi_X$ because IGG=Z$_2$.
So the gauge transformation does not change $\phi_{T_n} ~ (n=1,2,3)$.
As for $\phi_{\Pi_1}$, if $n_5 = 1$, $n_5 \pi r_2$ in (\ref{eq:PSG_Pi_1_eq_1}) is removed under the transformation.
Combining the above results with (\ref{eq:PSG_T_1}), (\ref{eq:PSG_T_2}), (\ref{eq:PSG_T_3}),
the PSG for $\{ T_1, T_2, T_3, \Pi_1 \}$ is described as follows:
\begin{eqnarray}
 \phi_{T_1} (r_1,r_2,r_3)_p 
 &=& 0,
 \nonumber\\
 \phi_{T_2} (r_1,r_2,r_3)_p 
 &=& n_1 \pi r_1,
 \nonumber\\
 \phi_{T_3} (r_1,r_2,r_3)_p 
 &=& n_2 \pi r_2 + n_3 \pi r_1,
 \nonumber\\
 \phi_{\Pi_1} (r_1,r_2,r_3)_p
 &=&
 \phi_{\Pi_1} (0,0,0)_p
 +
 n_6 \pi r_3
 \nonumber\\
 &-&
 \frac{1}{2} n_1 \pi r_2 (r_2-1)
 -
 n_3 \pi r_2 r_3,
 \nonumber
\end{eqnarray}
where
\begin{align}
 2 \phi_{\Pi_1} (0,0,0)_p = n_7 \pi.
 \nonumber
\end{align}

\subsection{$\phi_{\Pi_2} (r_1,r_2,r_3)_p$}
Conducting similar calculations as above for (\ref{eq:constraints_Pi_2})
leads to the following equations:
\begin{eqnarray}
 &&
 \phi_{\Pi_2} (r_1,r_2,r_3)_p 
 =
 \phi_{\Pi_2} (0,0,0)_p
 \nonumber\\
 &&
 ~~~~~~~~~~~~~~~~~~~~
 + 
 n_8 \pi r_1 
 +
 n_9 \pi r_2
 +
 n_{10} \pi r_3,
 \label{eq:PSG_Pi_2_eq_1}
 \\
 &&
 2 \phi_{\Pi_2} (0,0,0)_p = (n_{10} \cdot \delta_{p,B} + n_{11}) \pi,
 \label{eq:PSG_Pi_2_eq_2}
 \\
 &&
 n_3 = n_6 = n_8 = n_{12} = 0.
 \label{eq:PSG_Pi_2_eq_3}
\end{eqnarray}
(\ref{eq:PSG_Pi_2_eq_1}) is determined by (\ref{eq:constraints_Pi_2}a,b,c),
(\ref{eq:PSG_Pi_2_eq_2}) by (\ref{eq:constraints_Pi_2}d), and
(\ref{eq:PSG_Pi_2_eq_3}) by (\ref{eq:constraints_Pi_2}e).
Using this result,
the PSG for $\{ T_1, T_2, T_3, \Pi_1, \Pi_2 \}$ is given by
\begin{eqnarray}
 \phi_{T_1} (r_1,r_2,r_3)_p 
 &=& 0,
 \nonumber\\
 \phi_{T_2} (r_1,r_2,r_3)_p 
 &=& n_1 \pi r_1,
 \nonumber\\
 \phi_{T_3} (r_1,r_2,r_3)_p 
 &=& n_2 \pi r_2,
 \nonumber\\
 \phi_{\Pi_1} (r_1,r_2,r_3)_p
 &=&
 \phi_{\Pi_1} (0,0,0)_p
 -
 \frac{1}{2} n_1 \pi r_2 (r_2-1),
 \nonumber\\
 \phi_{\Pi_2} (r_1,r_2,r_3)_p
 &=&
 \phi_{\Pi_2} (0,0,0)_p
 +
 n_9 \pi r_2 + n_{10} \pi r_3,
 \nonumber
\end{eqnarray}
where
\begin{eqnarray}
 2 \phi_{\Pi_1} (0,0,0)_p &=& n_7 \pi,
 \nonumber\\
 2 \phi_{\Pi_2} (0,0,0)_p &=& ( n_{10} \cdot \delta_{p,B} + n_{11} ) \pi .
 \nonumber
\end{eqnarray}

\subsection{$\phi_{\Xi} (r_1,r_2,r_3)_p$}
Repeating the same procedure for (\ref{eq:constraints_Xi}),
we obtain
\begin{eqnarray}
 &&
 \phi_{\Xi} (r_1,r_2,r_3)_p
 =
 \phi_{\Xi} (0,0,0)_p
 \nonumber\\
 &&
 ~~~~~~~~~~~~~~~~~~
 +
 n_{13} \pi r_1
 +
 n_{14} \pi r_2
 +
 n_{15} \pi r_3,
 \label{eq:PSG_Xi_eq_1}
 \\
 &&
 \phi_{\Xi} (0,0,0)_p + \phi_{\Xi} (0,0,0)_{\bar{p}} =n_{16} \pi,
 \label{eq:PSG_Xi_eq_2}
\end{eqnarray}
\begin{subequations}
\begin{eqnarray}
 &&
 n_1=n_{13},
 \\
 &&
 \phi_{\Pi_1} (0,0,0)_B = \phi_{\Pi_1} (0,0,0)_{\bar{p}} + n_{17} \pi,
\end{eqnarray}
\label{eq:PSG_Xi_eq_3}
\end{subequations}
and
\begin{subequations}
\begin{eqnarray}
 &&
 n_{2}=n_{10}=n_{15}=0,
 \\
 &&
 \phi_{\Pi_2} (0,0,0)_B = \phi_{\Pi_2} (0,0,0)_{A} + n_{18} \pi.
\end{eqnarray}
\label{eq:PSG_Xi_eq_4}
\end{subequations}
The first equation (\ref{eq:PSG_Xi_eq_1}) is determined by (\ref{eq:constraints_Xi}a,b,c),
(\ref{eq:PSG_Xi_eq_2}) by (\ref{eq:constraints_Xi}d),
(\ref{eq:PSG_Xi_eq_3}) by (\ref{eq:constraints_Xi}e), and
(\ref{eq:PSG_Xi_eq_4}) by (\ref{eq:constraints_Xi}f).
Now we have the PSG for $\{ T_1, T_2, T_3, \Pi_1, \Pi_2, \Xi \}$,
which is described by
\begin{eqnarray}
 \phi_{T_1,3} (r_1,r_2,r_3)_p 
 &=& 0,
 \nonumber\\
 \phi_{T_2} (r_1,r_2,r_3)_p 
 &=& n_1 \pi r_1,
 \nonumber\\
 \phi_{\Pi_1} (r_1,r_2,r_3)_p
 &=&
 \phi_{\Pi_1} (0,0,0)_p 
 -
 \frac{1}{2} n_1 \pi r_2 (r_2-1),
 \nonumber\\
 \phi_{\Pi_2} (r_1,r_2,r_3)_p
 &=&
 \phi_{\Pi_2} (0,0,0)_p,
 +
 n_9 \pi r_2,
 \nonumber\\
 \phi_{\Xi} (r_1,r_2,r_3)_p
 &=&
 \phi_{\Xi} (0,0,0)_p,
 +
 n_1 \pi r_1
 +
 n_{14} \pi r_2,
 \nonumber
\end{eqnarray}
where
\begin{eqnarray}
 &&
 2 \phi_{\Pi_1} (0,0,0)_p = n_7 \pi,
 \nonumber\\
 &&
 2 \phi_{\Pi_2} (0,0,0)_p = n_{11} \pi,
 \nonumber\\
 &&
 \phi_{\Xi} (0,0,0)_{{A}} + \phi_{\Xi} (0,0,0)_{{B}} = n_{16} \pi,
 \nonumber\\
 &&
 \phi_{\Pi_1} (0,0,0)_B = \phi_{\Pi_1} (0,0,0)_{A} + n_{17} \pi. 
 \nonumber\\
 &&
 \phi_{\Pi_2} (0,0,0)_B = \phi_{\Pi_2} (0,0,0)_{A} + n_{18} \pi. 
 \nonumber
\end{eqnarray}

There are eight integer variables in the PSG obtained above.
However, 
in determining possible mean-field Hamiltonians,
most of those variables can be eliminated 
by using symmetries.
To be specific, we assume
\begin{subequations}
\begin{eqnarray}
 &&
 \eta_{(0,0,0)_A \rightarrow (1,0,0)_A} \ne 0,
 \\
 &&
 \eta_{(0,0,0)_A \rightarrow (0,1,0)_A} \ne 0,
 \\
 &&
 \eta_{(0,0,0)_A \rightarrow (0,0,0)_B} \ne 0,
 \\
 &&
 \eta_{(0,0,0)_A \rightarrow (1,1,0)_B} \ne 0.
\end{eqnarray}
\label{eq:four-indep-MF-parameters}
\end{subequations}
It must be noted that the above four mean-field parameters are independent
because they can not be transformed to each other via any symmetry operation in SG$_1$.
(This fact will be used when we construct the mean-field ans\"{a}tze in Appendix \ref{appendix:SG2_ansatze}).
Applying the equation (\ref{eq:symmetry-related-bonds}) to the above nonzero mean-field parameters,
we can find additional constraints on the eight integer variables and then obtain (\ref{eq:PSG_for_SG_1}).
First, we simplify $\phi_{\Pi}$.
Noting that the link $(0,0,0)_A \rightarrow (0,0,0)_B$ does not move under the operation $\Pi_1$ and
using the equation (\ref{eq:symmetry-related-bonds}) with $X=\Pi_1$, $i=(0,0,0)_A$, $j=(0,0,0)_B$,
we have 
\begin{eqnarray}
 &&
 \phi_{\Pi_1} (0,0,0)_A + \phi_{\Pi_1} (0,0,0)_B = 0.
 \label{eq:addtional_constraints_1}
\end{eqnarray}
Next, 
we note that 
for both of the sublattice $p=A,B$, 
\begin{eqnarray}
 &&
 (-1,0,0)_p \rightarrow (0,0,0)_p
 \xrightarrow{T_1}
 (0,0,0)_p \rightarrow (1,0,0)_p
 \nonumber\\
 &\xrightarrow{\Pi_1}&
 (0,0,0)_p \rightarrow (-1,0,0)_p
 =
 -(-1,0,0)_p \rightarrow (0,0,0)_p.
 \nonumber
\end{eqnarray}
In this case, the equation (\ref{eq:symmetry-related-bonds}) is written as follows:
\begin{eqnarray}
 &&
 -\eta_{{(-1,0,0)_p \rightarrow (0,0,0)_p}} 
 \nonumber\\
 &=&
 \eta_{ (0,0,0)_p \rightarrow (1,0,0)_p }
 \cdot
 e^{ - i \phi_{\Pi_1} [\Pi_1(0,0,0)_p] - i \phi_{\Pi_1} [\Pi_1(1,0,0)_p] }
 \nonumber\\
 &=&
 \eta_{ (-1,0,0)_p \rightarrow (0,0,0)_p }
 \cdot
 e^{ - i \phi_{\Pi_1} [\Pi_1(0,0,0)_p] - i \phi_{\Pi_1} [\Pi_1(1,0,0)_p] },
 \nonumber
\end{eqnarray}
where we used $\phi_{T_1}=0$.
The above equation is reduced to
\begin{eqnarray}
 \phi_{\Pi_1} [\Pi_1(0,0,0)_p] + \phi_{\Pi_1} [\Pi_1(1,0,0)_p] = \pi.
 \nonumber
\end{eqnarray}
Simplifying the above equation, we get the condition
\begin{eqnarray}
 &&
 2 \phi_{\Pi_1} (0,0,0)_p = \pi ~ (p=A,B)
 \label{eq:addtional_constraints_2}
 \\
 & \Rightarrow &
 n_7=1.
 \nonumber
\end{eqnarray}
Subtracting (\ref{eq:addtional_constraints_1}) from (\ref{eq:addtional_constraints_2}),
we find one more condition:
\begin{eqnarray}
 &&
 \phi_{\Pi_1} (0,0,0)_B - \phi_{\Pi_1} (0,0,0)_{A} = \pi
 \label{eq:addtional_constraints_3}
 \\
 & \Rightarrow &
 n_{17}=1.
 \nonumber
\end{eqnarray}
Then, $\phi_{\Pi_1}(0,0,0)_p$ is determined by (\ref{eq:addtional_constraints_1}), (\ref{eq:addtional_constraints_2}), and (\ref{eq:addtional_constraints_3}):
\begin{eqnarray}
 \phi_{\Pi_1} (0,0,0)_p 
 = 
 \left( \frac{1}{2} + m_1 + \delta_{p,B} \right) \pi,
 \nonumber
\end{eqnarray}
where $m_1 \in \{0,1\}$.
Ignoring the constant $m_1 \pi$ (because IGG=Z$_2$),
we obtain (\ref{eq:PSG_for_SG_1_Pi_1}).

Next, to simplify $\phi_{\Pi_2}$,
we notice that the link $(0,0,0)_A \rightarrow (1,0,0)_A$ is invariant under $\Pi_2$.
From this fact,
\begin{eqnarray}
 &&
 \phi_{\Pi_2} (0,0,0)_A + \phi_{\Pi_2} (1,0,0)_A = 0
 \nonumber\\
 & \Rightarrow &
 2 \phi_{\Pi_2} (0,0,0)_A = 0
 \nonumber\\
 & \Rightarrow &
 n_{11}=0,
 \nonumber
\end{eqnarray}
Then, we have $2 \phi_{\Pi_2} (0,0,0)_p = n_{11} \pi = 0 ~ (p=A,B)$,
from which $\phi_{\Pi_2} (0,0,0)_p$ can be written as
\begin{eqnarray}
 \phi_{\Pi_2} (0,0,0)_p = (m_2 + m_3 \cdot \delta_{p,B}) \pi,
 \nonumber
\end{eqnarray}
where $m_2, m_3 \in \{ 0, 1 \}$.
Considering another link invariant under $\Pi_2$, $(0,0,0)_A \rightarrow (0,1,0)_A$,
we get additional constraint:
\begin{eqnarray}
 &&
 \phi_{\Pi_2} (0,0,0)_A + \phi_{\Pi_2} (0,1,0)_A = 0
 \nonumber\\
 & \Rightarrow &
 2 \phi_{\Pi_2} (0,0,0)_A + n_9 \pi = 0
 \nonumber\\
 & \Rightarrow &
 n_{9}=0.
 \nonumber
\end{eqnarray}
With above conditions,
$\phi_{\Pi_2}$ can be written as
\begin{eqnarray}
 \phi_{\Pi_2}(r_1,r_2,r_3)_p
 &=&
 \phi_{\Pi_2}(0,0,0)_p
 \nonumber\\
 &=&
 (m_2 + m_3 \cdot \delta_{p,B}) \pi.
 \label{eq:PSG_for_SG_1_Pi_2_intermediate}
\end{eqnarray}
$\phi_{\Pi_2}$ can be simplified further by taking following gauge transformation $G_5=e^{i\phi_5}$:
\begin{eqnarray}
 \phi_5 (r_1,r_2,r_3)_p 
 = 
 \pi r_3.
 \nonumber
\end{eqnarray}
Under the transformation,
only $\phi_{\Pi_2}$ is affected as follows.
\begin{eqnarray}
 \phi_{\Pi_2} (r_1,r_2,r_3)_p 
 \rightarrow 
 \phi_{\Pi_2} (r_1,r_2,r_3)_p + \pi \cdot \delta_{p,B}.
 \nonumber
\end{eqnarray}
With this transformation, 
$\delta_{p,B} \cdot m_3  \pi$ in (\ref{eq:PSG_for_SG_1_Pi_2_intermediate}) can be eliminated.
Ignoring $m_2 \pi$ in the equation as well (IGG=Z$_2$),
the equation is reduced to (\ref{eq:PSG_for_SG_1_Pi_2}).

Finally, we simplify $\phi_{\Xi}$.
Let us consider the link $(0,0,0)_A \rightarrow (0,0,0)_B$.
\begin{eqnarray}
 (0,0,0)_A \rightarrow (0,0,0)_B 
 & \xrightarrow{\Xi} &
 (0,0,0)_B \rightarrow (0,0,0)_A 
 \nonumber\\
 &=& 
 -
 (0,0,0)_A \rightarrow (0,0,0)_B.
 \nonumber
\end{eqnarray}
From this, we obtain
\begin{eqnarray}
 &&
 \phi_{\Xi} (0,0,0)_{{A}} + \phi_{\Xi} (0,0,0)_{{B}} = \pi
 \nonumber\\
 & \Rightarrow &
 n_{16} = 1.
 \label{eq:addtional_constraints_6}
\end{eqnarray}
We also note that
\begin{eqnarray}
 &&
 (-1,-1,0)_A \rightarrow (0,0,0)_B
 \xrightarrow{T_1 T_2} 
 (0,0,0)_A \rightarrow (1,1,0)_B
 \nonumber\\
 & \xrightarrow{\Xi} &
 (0,0,0)_B \rightarrow (-1,-1,0)_A
 =
 - 
 (-1,-1,0)_A \rightarrow (0,0,0)_B .
 \nonumber
\end{eqnarray}
It follows that
\begin{eqnarray}
 &&
 \left[ \phi_{T_2} (-1,-1,0)_A + \phi_{T_2} (0,0,0)_B \right]
 \nonumber\\
 &+&
 \left[ \phi_{\Xi} (0,0,0)_A + \phi_{\Xi} (1,1,0)_B \right]
 =
 \pi
 \nonumber\\
 & \Rightarrow &
 \phi_{\Xi} (0,0,0)_A + \phi_{\Xi} (0,0,0)_B + n_{14} \pi = \pi
 \nonumber\\
 & \Rightarrow &
 n_{14} = 0.
 \label{eq:addtional_constraints_7}
\end{eqnarray}
From the conditions (\ref{eq:addtional_constraints_6}) and (\ref{eq:addtional_constraints_7}),
$\phi_{\Xi}$ can be written as follows:
\begin{eqnarray}
 \phi_{\Xi} (r_1,r_2,r_2)_p 
 &=& 
 \phi_{\Xi} (0,0,0)_p + n_1 \pi r_1,
 \nonumber
\end{eqnarray}
where
\begin{eqnarray}
 \phi_{\Xi} (0,0,0)_p
 =
 \left\{
 \begin{array}{cc}
 \phi_0 & (p=A)
 \\
 \pi - \phi_0 & (p=B) 
 \end{array}
 \right. .
 \nonumber
\end{eqnarray}
The dependence of $\phi_{\Xi}$ on the sublattice can be eliminated by the gauge transformation $G_6=e^{i\phi_{G_6}}$,
where
\begin{align}
 \phi_{6} (r_1,r_2,r_3)_p
 =
 \left\{
 \begin{array}{cc}
  \phi_1 & (p=A)
  \\
  -\phi_1 & (p=B)
 \end{array}
 \right. .
 \nonumber
 \nonumber
\end{align}
Under the transformation,
only $\phi_{\Xi}$ changes, in the following way:
\begin{align}
 \phi_{\Xi} (r_1,r_2,r_3)_p 
 \rightarrow
 \phi_{\Xi} (r_1,r_2,r_3)_p + 2 \phi_1 \cdot (-1)^{\delta_{p,B}}.
 \nonumber
\end{align}
By choosing $2 \phi_1 = \pi/2 - \phi_0$,
we obtain 
\begin{align}
 \phi_{\Xi} (r_1,r_2,r_3)_p = \frac{\pi}{2} + n_1 \pi r_1,
 \nonumber
\end{align}
which is same as (\ref{eq:PSG_for_SG_1_Xi}).

\subsection{$\phi_{R} (r_1,r_2,r_3)_p${\label{appendix:PSG-construction-2-rotation}}}
We now consider adding the rotation $R$ to the symmetry group, in order to
find the PSGs for SG$_2$.
We obtain following conditions from (\ref{eq:constraints_R}a,b,c).
\begin{eqnarray}
 &&
 \phi_{R} (r_1,r_2,r_3)_p
 \nonumber\\
 &=&
 \phi_{R} (0,0,0)_p
 +
 n_{19} \pi r_1
 +
 n_{20} \pi r_2 
 +
 n_{21} \pi r_3
 \nonumber\\
 &+&
 n_1 \pi r_1 r_2 - \frac{1}{2} n_1 \pi r_1 (r_1 -1). 
 \label{eq:additional_condition_phi_R_0}
\end{eqnarray}
From (\ref{eq:constraints_R}d),
we get two conditions:
\begin{subequations}
\begin{eqnarray}
 &&
 n_{21}=0,
 \\
 &&
 3 \phi_R (0,0,0)_p = n_{22} \pi - (n_1+n_{19}) \pi \cdot \delta_{p,B}.
\end{eqnarray}
\label{eq:additional_condition_phi_R_1}
\end{subequations}
By using the constraint from (\ref{eq:constraints_R}e),
\begin{subequations}
\begin{eqnarray}
 &&
 n_1=n_{19}+n_{20},
 \\
 &&
 2\phi_{R}(0,0,0)_p=n_{23} \pi.
\end{eqnarray}
\label{eq:additional_condition_phi_R_2}
\end{subequations}
Using the conditions (\ref{eq:additional_condition_phi_R_1}) and (\ref{eq:additional_condition_phi_R_2}),
we can simplify (\ref{eq:additional_condition_phi_R_0}) as follows.
\begin{eqnarray}
 \phi_{R} (r_1,r_2,r_3)_p
 &=&
 \phi_{R} (0,0,0)_p
 +
 n_{19} \pi r_1
 +
 n_{20} \pi r_2 
 \nonumber\\
 &+&
 n_1 \pi r_1 r_2 - \frac{1}{2} n_1 \pi r_1 (r_1 -1),
 \label{eq:phi_R_prefinal_1}
\end{eqnarray}
where
\begin{eqnarray}
 &&
 n_1=n_{19}+n_{20},
 \nonumber\\
 &&
 \phi_R (0,0,0)_p = ( n_{22} + n_{23} )\pi + n_{20} \pi \cdot \delta_{p,B}.
 \label{eq:phi_R_prefinal_2}
\end{eqnarray}
The equation (\ref{eq:constraints_R}f,g) does not give a new condition on the integer variables.

The above expression for $\phi_{R}$ can be further simplified by using rotational symmetries for certain links.
First, we note that
\begin{eqnarray}
 (0,0,0)_A \rightarrow (1,1,0)_B
 & \xrightarrow{R} &
 (0,0,0)_A \rightarrow (0,1,0)_B
 \nonumber\\
 & \xrightarrow{\Pi_1} &
 (0,0,0)_A \rightarrow (1,1,0)_B.
 \nonumber
\end{eqnarray}
Applying (\ref{eq:symmetry-related-bonds}) to the above relation,
\begin{eqnarray}
 &&
 \phi_R (0,0,0)_A + \phi_R (0,0,0)_B = 0
 \nonumber\\
 & \Rightarrow &
 n_{20}=0
 \Rightarrow
 n_{19}=n_1.
 \label{eq:last_condition_1}
\end{eqnarray}
The last equation is obtained from the fact that $n_1=n_{19}+n_{20}$.
Next, we consider following symmetry:
\begin{eqnarray}
 &&
 (1,0,0)_A \rightarrow (1,1,0)_A
 \xrightarrow{R}
 (0,1,0)_A \rightarrow (-1,0,0)_A
 \nonumber\\
 & \xrightarrow{\Pi_1} &
 (1,1,0)_A \rightarrow (1,0,0)_A
 =
 -(1,0,0)_A \rightarrow (1,1,0)_A.
 \nonumber
\end{eqnarray}
For this case,
(\ref{eq:symmetry-related-bonds}) gives us 
\begin{eqnarray}
 &&
 \left[ \phi_R (1,0,0)_A + \phi_R (1,1,0)_A \right]
 \nonumber\\
 &+&
 \left[ \phi_{\pi_1} (0,1,0)_A + \phi_{\pi_1} (-1,0,0)_A \right]
 =
 \pi
 \nonumber\\
 & \Rightarrow &
 n_1=0.
 \label{eq:last_condition_2}
\end{eqnarray}
Plugging the conditions (\ref{eq:last_condition_1}) and (\ref{eq:last_condition_2}) 
into (\ref{eq:phi_R_prefinal_1}) and (\ref{eq:phi_R_prefinal_2}),
we have following expression for $\phi_{R}$:
\begin{align}
 \phi_{R} (r_1,r_2,r_3)_p
 = 
 (n_{22}+n_{23}) \pi.
 \label{eq:phi_R_prefinal_3}
\end{align}
It must be remembered that 
the site-independent constant $(n_{22}+n_{23}) \pi$ in the equation can be ignored due to IGG=Z$_2$.
The conditions (\ref{eq:last_condition_2}) and (\ref{eq:phi_R_prefinal_3}) lead to (\ref{eq:PSG_for_SG_2}), the PSG for the SG$_2$.

\section{Brief sketch of the derivation of mean-field ans\"{a}tze\label{appendix:SG2_ansatze}}

In this section,
we construct the ans\"{a}tze of the SG$_1$.
To find out the possible ans\"{a}tze for a given symmetry group,
we must classify all possible projective symmetry groups.
In the PSG for the SG$_1$ (\ref{eq:PSG_for_SG_1}),
there is only one integer variable: $n_1=0,1$.
Different values of $n_1$ correspond to distinct PSGs.
However, there is another factor to determine the PSGs in addition to $n_1$,
which is the relative sign between the four independent mean-field parameters (\ref{eq:four-indep-MF-parameters}).
We assume that the mean-field parameters is real-valued to preserve the time reversal symmetry.
Then, they can be positive or negative.
However, we can fix three of those parameters to be positive in a certain orientation by employing several gauge transformations below.

First, the mean-field parameter $\eta_{(0,0,0)_A \rightarrow (1,0,0)_A}$ can be fixed to be positive via the transformation $G_7=e^{i\phi_7}$, 
where
\begin{eqnarray}
 \phi_7 (r_1,r_2,r_3)_p = \textup{constant.}
 \label{eq:gauge-transf-7}
\end{eqnarray}
Next, we can render the parameter $\eta_{(0,0,0)_A \rightarrow (0,1,0)_A}$ positive by using the transformation $G_8=e^{i\phi_8}$, 
where
\begin{eqnarray}
 \phi_8 (r_1,r_2,r_3)_p = \pi r_2.
 \label{eq:gauge-transf-8}
\end{eqnarray}
Finally, the parameter $\eta_{(0,0,0)_A \rightarrow (0,0,0)_B}$ becomes positive in the help of the transformation $G_9=e^{i\phi_9}$ with
\begin{eqnarray}
 \phi_9 (r_1,r_2,r_3)_p = \pi \cdot \delta_{p,A}.
 \label{eq:gauge-transf-9}
\end{eqnarray}
Under the transformations (\ref{eq:gauge-transf-7}), (\ref{eq:gauge-transf-8}), (\ref{eq:gauge-transf-9}),
the PSGs (\ref{eq:PSG_for_SG_1}) and (\ref{eq:PSG_for_SG_2}) are invariant up to a constant $\pi$,
which can be ignored due to IGG=Z$_2$.
Therefore, the parameter $\eta_{(0,0,0)_A \rightarrow (1,1,0)_B}$ is only free to change its sign.
Now, we define positive mean-field parameters ($\eta_{1\alpha}$, $\eta_{1\beta}$, $\eta_{2\alpha}$, $\eta_{2\beta}$) as follows:
\begin{subequations}
\begin{eqnarray}
 &&
 \eta_{(0,0,0)_A \rightarrow (1,0,0)_A} \equiv \eta_{1\alpha},
 \\
 &&
 \eta_{(0,0,0)_A \rightarrow (0,1,0)_A} \equiv \eta_{1\beta},
 \\
 &&
 \eta_{(0,0,0)_A \rightarrow (0,0,0)_B} \equiv \eta_{2\alpha},
 \\
 &&
 \eta_{(0,0,0)_A \rightarrow (1,1,0)_B} \equiv \eta_{2\beta} \cdot (-1)^{m},
\end{eqnarray}
\label{eq:four-indep-MF-parameters-2}
\end{subequations}
where $m=0,1$.
The integer variable $m$ is the other factor to determine the ans\"{a}tze in addition to $n_1$.
Therefore, there are four distinct ans\"atze corresponding to four different combinations of $n_1$ and $m$.
The four ans\"atze are classified in Table \ref{tab:ansatze of SG_1}.

Now we construct the mean-field ans\"atze for the symmetry group SG$_1$.
The ans\"{a}tze are constructed by using (\ref{eq:symmetry-related-bonds}) and (\ref{eq:PSG_for_SG_1}).
This generates the values of all symmetry-related mean-field parameters from a single bond,
through the application of elements of the PSG.

\subsection{$(0,\pi)$-flux and $(0,0)$-flux ans\"atze}

First, we consider the cases of $n_1=0$.
If we set $X=T_n ~ (n=1,2,3)$ in (\ref{eq:symmetry-related-bonds}) and use (\ref{eq:PSG_for_SG_1_T_1}) and (\ref{eq:PSG_for_SG_1_T_23}) with $n_1=0$,
then we have
\begin{eqnarray}
 \eta_{T_n(i) T_n(j)} = \eta_{ij}.
 \nonumber
\end{eqnarray}
This means that the mean-field ans\"{a}tze with $n_1=0$ are translationally invariant.
Therefore, it is enough to specify the mean-field structure within a unit cell for the ans\"atze.

\begin{figure}
 \centering
 \includegraphics[width=0.8\linewidth]{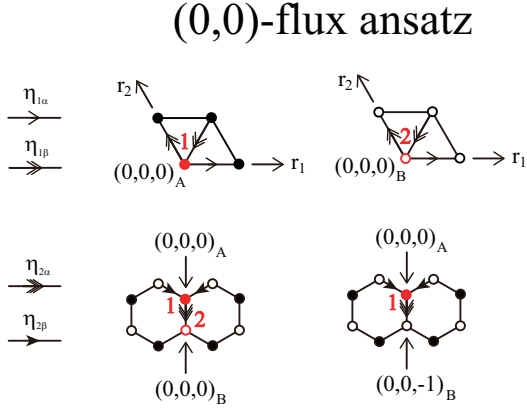}
 \caption{
 (Color online) 
 (0,0)-flux ansatz.
 The figure shows the two sites (red and numbered) and the twelve mean-fields (arrowed links) in a unit cell.
 The full mean-field configuration is obtained via translations along the three lattice vectors $\{ {\bf R}_1 , {\bf R}_2 , {\bf R}_3 \}$.
 Switching the positive orientation of $\eta_{2\beta}$ in the figure leads to the (0,$\pi$)-flux ansatz.
 \label{fig:zero-zero_ansatz}}
\end{figure}

Considering the unit cell denoted with the ellipse in Fig. \ref{fig:coordinate_system},
the twelve links consist of six intra-layer links (\ref{eq:intralayer_link})
and six inter-layer links (\ref{eq:interlayer_link}).  
\begin{subequations}
\label{eq:intralayer_link}
\begin{eqnarray}
 &&
 (0,0,0)_A \rightarrow (1,0,0)_A,
 \label{eq:intralayer_link_1}
 \\
 &&
 (0,0,0)_A \rightarrow (0,1,0)_A,
 \label{eq:intralayer_link_2}
 \\
 &&
 (0,0,0)_A \rightarrow (1,1,0)_A,
 \label{eq:intralayer_link_3}
 \\
  &&
 (0,0,0)_B \rightarrow (-1,0,0)_B,
 \label{eq:intralayer_link_4}
 \\
 &&
 (0,0,0)_B \rightarrow (0,-1,0)_B,
 \label{eq:intralayer_link_5}
 \\
 &&
 (0,0,0)_B \rightarrow (-1,-1,0)_B,
 \label{eq:intralayer_link_6}
\end{eqnarray}
\end{subequations}
\begin{subequations}
\label{eq:interlayer_link}
\begin{eqnarray}
 &&
 (0,0,0)_A \rightarrow (0,0,0)_B,
 \label{eq:interlayer_link_1}
 \\
 &&
 (0,0,0)_A \rightarrow (1,1,0)_B,
 \label{eq:interlayer_link_2}
 \\
 &&
 (0,0,0)_A \rightarrow (0,1,0)_B,
 \label{eq:interlayer_link_3}
 \\
  &&
 (0,0,0)_A \rightarrow (0,0,-1)_B,
 \label{eq:interlayer_link_4}
 \\
 &&
 (0,0,0)_A \rightarrow (1,1,-1)_B,
 \label{eq:interlayer_link_5}
 \\
 &&
 (0,0,0)_A \rightarrow (0,1,-1)_B.
 \label{eq:interlayer_link_6}
\end{eqnarray}
\end{subequations}
We fix mean-field parameters at four links, 
(\ref{eq:intralayer_link_1}), 
(\ref{eq:intralayer_link_2}), 
(\ref{eq:interlayer_link_1}), 
and
(\ref{eq:interlayer_link_2})
, as mentioned in (\ref{eq:four-indep-MF-parameters-2}).
The links (\ref{eq:intralayer_link}) and (\ref{eq:interlayer_link}) within the unit cell 
are connected by symmetry operations of SG$_1$ in the following way:
\begin{eqnarray}
 &\boxed{(\textup{\ref{eq:intralayer_link_1}})}& ~~~~~~ \boxed{(\textup{\ref{eq:intralayer_link_2}})} \xrightarrow{\Pi_1} (\textup{\ref{eq:intralayer_link_3}})
 \nonumber\\
 &{\Xi} \downarrow ~ &  ~~~~~~~  {\Xi} \downarrow  ~~~~~~~~~  {\Xi} \downarrow
 \nonumber\\
 &(\textup{\ref{eq:intralayer_link_4}})& ~~~~~~~ (\textup{\ref{eq:intralayer_link_5}}) ~ \xrightarrow{\Pi_1} (\textup{\ref{eq:intralayer_link_6}}),
 \nonumber\\
 \nonumber\\
 &\boxed{(\textup{\ref{eq:interlayer_link_1}})}& ~~~~~~ \boxed{(\textup{\ref{eq:interlayer_link_2}})} \xrightarrow{\Pi_1} (\textup{\ref{eq:interlayer_link_3}})
 \nonumber\\
 & {\Pi_2} \downarrow ~~ &  ~~~~~~  {\Pi_2} \downarrow  ~~~~~~~  {\Pi_2} \downarrow
 \nonumber\\
 &(\textup{\ref{eq:interlayer_link_4}})& ~~~~~~~ (\textup{\ref{eq:interlayer_link_5}}) ~ \xrightarrow{\Pi_1} (\textup{\ref{eq:interlayer_link_6}}).
 \label{eq:SG1_symmetry_relationship}
\end{eqnarray}
In the above diagram,
the boxes denote the links where the mean-field parameter is fixed.
If we apply (\ref{eq:symmetry-related-bonds}) to (\ref{eq:SG1_symmetry_relationship}),
we can determine the other eight mean-field parameters in the unit cell.
Their mean-field configurations in the unit cell are given as follows:
\begin{subequations}
\begin{eqnarray}
 \eta_{1\alpha}
 &=&
 \eta_{ (0,0,0)_A \rightarrow (1,0,0)_A } 
 =
 \eta_{ (-1,0,0)_B \rightarrow (0,0,0)_B }
 ,
 \nonumber\\
 \eta_{1\beta}
 &=&
 \eta_{ (0,0,0)_A \rightarrow (0,1,0)_A } 
 =
 \eta_{ {{(1,1,0)_A \rightarrow (0,0,0)_A}} }
 \nonumber\\
 &=&
 \eta_{ (0,-1,0)_B \rightarrow (0,0,0)_B } 
 =
 \eta_{ {{(0,0,0)_B \rightarrow (-1,-1,0)_B}} }
 ,
 \nonumber\\
 \eta_{2\alpha}
 &=&
 \eta_{ (0,0,0)_A \rightarrow (0,0,0)_B } 
 =
 \eta_{ (0,0,0)_A \rightarrow (0,0,-1)_B }
 ,
 \nonumber\\
 (-1)^{m}
 \cdot
 \eta_{2\beta}
 &=&
 \eta_{ (0,0,0)_A \rightarrow (1,1,0)_B } 
 =
 \eta_{ (0,0,0)_A \rightarrow (1,1,-1)_B } 
 \nonumber\\
 &=&
 \eta_{ (0,0,0)_A \rightarrow (0,1,0)_B } 
 =
 \eta_{ (0,0,0)_A \rightarrow (0,1,-1)_B }
 .
 \nonumber
\end{eqnarray}
\end{subequations}
Here, the case of $m=1$ corresponds to (0,0)-flux ansatz and $m=0$ does to (0,$\pi$)-flux ansatz.
The mean-field configuration in a unit cell of the (0,0)-flux ansatz is depicted in Fig. \ref{fig:zero-zero_ansatz}.
Switching the positive orientation of $\eta_{2\beta}$ in the figure leads to the (0,$\pi$)-flux ansatz.

\subsection{$(\pi,0)$-flux and $(\pi,\pi)$-flux ans\"atze}

\begin{figure}
 \centering
 \includegraphics[width=0.9\linewidth]{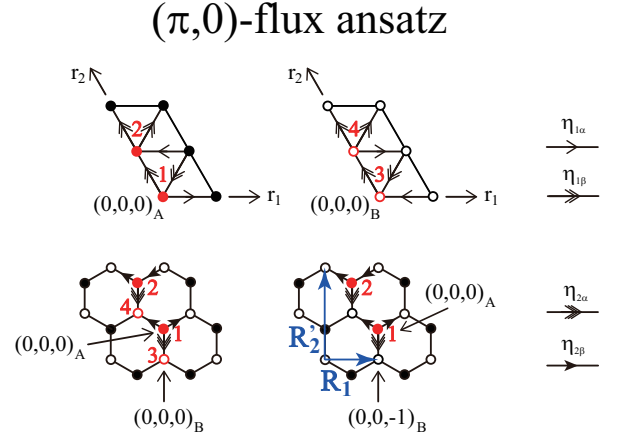}
 \caption{
 (Color online)
 ($\pi$,0)-flux ansatz.
 The figure shows the four sites (red and numbered) and the twenty four mean-fields (arrowed links) in a unit cell.
 The ansatz has the orthorhombic Bravais lattice structure.
 The lattice vectors of the orthorhombic lattice are given by $\{ {\bf R}_1, {\bf R}'_2, {\bf R}_3 \}$,
 where  ${\bf R}'_2={\bf R}_1+2{\bf R}_2$.
 The full mean-field configuration is obtained via translations along the three lattice vectors.
 Switching the positive orientation of $\eta_{2\beta}$ in the figure leads to the ($\pi$,$\pi$)-flux ansatz.
 \label{fig:pi-zero_ansatz}}
\end{figure}

Repeating the same procedure above for the cases of $n_1=1$,
we can obtain following ans\"atze.
\begin{subequations}
\begin{eqnarray}
 \eta_{1\alpha}
 & = &
 \eta_{ (0,0,0)_A \rightarrow (1,0,0)_A } 
 =
 \eta_{ (1,1,0)_A \rightarrow (0,1,0)_A } 
 \nonumber\\
 &=&
 \eta_{ (1,0,0)_B \rightarrow (0,0,0)_B }
 =
 \eta_{ (0,1,0)_B \rightarrow (1,1,0)_B }
 ,
 \nonumber\\
 \eta_{1\beta}
 & = &
 \eta_{ (0,0,0)_A \rightarrow (0,1,0)_A } 
 =
 \eta_{ (0,1,0)_A \rightarrow (0,2,0)_A } 
 \nonumber\\
 & = &
 \eta_{ (1,1,0)_A \rightarrow (0,0,0)_A } 
 =
 \eta_{ (0,1,0)_A \rightarrow (1,2,0)_A } 
 \nonumber\\
 & = &
 \eta_{ (0,0,0)_B \rightarrow (0,1,0)_B } 
 =
 \eta_{ (0,1,0)_B \rightarrow (0,2,0)_B } 
 \nonumber\\
 & = &
 \eta_{ (1,1,0)_B \rightarrow (0,0,0)_B } 
 =
 \eta_{ (0,1,0)_B \rightarrow (1,2,0)_B } 
 ,
 \nonumber\\
 \eta_{2\alpha}
 & = &
 \eta_{ (0,0,0)_A \rightarrow (0,0,0)_B } 
 =
 \eta_{ (0,1,0)_A \rightarrow (0,1,0)_B }  
 \nonumber\\
 & = &
 \eta_{ (0,0,0)_A \rightarrow (0,0,-1)_B }
 =
 \eta_{ (0,1,0)_A \rightarrow (0,1,-1)_B } 
 ,
 \nonumber\\
 (-1)^m
 \cdot
 \eta_{2\beta}
 & = &
 \eta_{ (0,0,0)_A \rightarrow (1,1,0)_B } 
 =
 \eta_{ (0,0,0)_A \rightarrow (0,1,0)_B } 
 \nonumber\\
 & = &
 \eta_{ (0,1,0)_A \rightarrow (0,2,0)_B } 
 =
 \eta_{ (1,2,0)_B \rightarrow (0,1,0)_A } 
 \nonumber\\
 & = &
 \eta_{ (0,0,0)_A \rightarrow (1,1,-1)_B } 
 =
 \eta_{ (0,0,0)_A \rightarrow (0,1,-1)_B } 
 \nonumber\\
 & = &
 \eta_{ (0,1,0)_A \rightarrow (0,2,-1)_B } 
 =
 \eta_{ (1,2,-1)_B \rightarrow (0,1,0)_A } 
 .
 \nonumber
\end{eqnarray}
\end{subequations}
The two ans\"atze have enlarged unit cell
because $\phi_{T_2}$ is site-dependent when $n_1=1$ [see (\ref{eq:PSG_for_SG_1_T_23})].
There are four sites and twenty four links in a unit cell.
The case of $m=0$ corresponds to ($\pi$,0)-flux ansatz and $m=1$ does to ($\pi$,$\pi$)-flux ansatz.
Fig. \ref{fig:pi-zero_ansatz} shows the mean-field configuration in a unit cell of the ($\pi$,0)-flux ansatz.
The configuration for the ($\pi$,$\pi$)-flux ansatz is obtained by switching the positive orientation of $\eta_{2\beta}$ in the figure.

\subsection{Symmetric ansatz}
The symmetric ansatz with the full symmetry group SG$_2$ can be constructed
by considering the constraints due to the rotational symmetry: 
$n_1=0$, $m=0$, and $\eta_{1\alpha}=\eta_{1\beta}$ and $\eta_{2\alpha}=\eta_{2\beta}$.
The condition $n_1=0$ was already found in Section \ref{appendix:PSG-construction-2-rotation}.
The rest of the conditions come from the fact that $\phi_R=0$ in (\ref{eq:PSG_for_SG_2}).
Therefore, the symmetric ansatz is a special case of the (0,$\pi$)-flux ansatz 
with $\eta_{1\alpha}=\eta_{1\beta}$ and $\eta_{2\alpha}=\eta_{2\beta}$.

\section{Mean-field Hamiltonians\label{appendix:mean-field Hamiltonians}}

In this appendix,
we provide the mean-field Hamiltonians of the ans\"atze obtained in previous section in momentum space.
For each ansatz, the mean-field Hamiltonian takes the following form:
\begin{eqnarray}
 H_{MF} 
 =
 N_{uc} \cdot n_s \cdot \epsilon_{0}
 + \sum_{\bf k} \Psi^{\dagger} ({\bf k}) {\bf D} ({\bf k}) \Psi ({\bf k}),
 \label{eq:H_MF_momentum_space}
\end{eqnarray}
where
\begin{eqnarray}
 \epsilon_{0} 
 &=&
 - \lambda ( \kappa + 1 ) 
 \nonumber\\
 && 
 + \sum_{n=1}^{2} J_n \left(\frac{1}{2} \left| \eta_{n\alpha} \right|^2  + \left| \eta_{n\beta} \right|^2 + \frac{3}{4} \kappa^2 \right),
 \nonumber
\end{eqnarray} 
\begin{eqnarray}
 \Psi ({\bf k}) 
 = 
 ( b_{1 \uparrow} ({\bf k}), 
   \cdots,
   b_{n_s \uparrow} ({\bf k}), 
   b_{1 \downarrow}^{\dagger} (-{\bf k}), 
  \cdots,
   b_{n_s \downarrow}^{\dagger} (-{\bf k})
 )^T,
 \nonumber
\end{eqnarray}
\begin{eqnarray}
 {\bf D}({\bf k})
 =
 \left(
 \begin{array}{c|c}
 \lambda \mathbb{I}_{n_s} & {\bf F}({\bf k})
 \\
 \hline
 {\bf F}^{\dagger}({\bf k}) & \lambda \mathbb{I}_{n_s}
 \end{array}
 \right).
 \nonumber
\end{eqnarray}
In the above expressions, 
$N_{uc}$ is the number of unit cells and $n_s$ is the number of sites in a unit cell.
The (0,0)-flux and (0,$\pi$)-flux ans\"atze have $n_s=2$
and 
the ($\pi$,0)-flux and ($\pi$,$\pi$)-flux ans\"atze have $n_s=4$.
$b_{n \mu} ({\bf k}) ~ (n=1,\cdots,n_s;\mu=\uparrow,\downarrow)$ is the boson operator in momentum space.
The subscript $n$ denotes the site in a unit cell and the site-numbering is shown in Fig. \ref{fig:zero-zero_ansatz} and \ref{fig:pi-zero_ansatz} (red numbers).
$\lambda$ is the Lagrange multiplier, which is taken to be uniform.
$\mathbb{I}_{n_s}$ is the $n_s \times n_s$ identity matrix.
The other $n_s \times n_s$ matrix ${\bf F}({\bf k})$ is defined as follows:
\begin{eqnarray}
 {\bf F}_{nn'}({\bf k})
 =
 &-&
 \frac{1}{2} 
 \sum_{ (l,l',{\bf R}) \in \textup{L} }
 J_{ll'}({\bf R}) \eta_{ll'}({\bf R}) 
 \cdot
 e^{i{\bf k}\cdot{\bf R}}
 \cdot
 \delta_{nl} \delta_{n'l'}
 \nonumber\\
 &+&
 \frac{1}{2} 
 \sum_{ (l,l',{\bf R}) \in \textup{L} }
 J_{ll'}({\bf R}) \eta_{ll'}({\bf R}) 
 \cdot
 e^{-i{\bf k}\cdot{\bf R}}
 \cdot
 \delta_{nl'} \delta_{n'l},
 \nonumber\\
 \label{eq:Hamiltonian-submatrix}
\end{eqnarray}
where the set $\textup{L}$ and $J_{ll'}({\bf R})$ and $\eta_{ll'}({\bf R})$ are listed in Table \ref{tab:MF-links-1} and \ref{tab:MF-links-2}.

\begin{table}
\begin{ruledtabular}
\begin{tabular}{ccccc}
$l$ & $l'$ & ${\bf R}$ & $J_{ll'}({\bf R})$ & $\eta_{ll'}({\bf R})$
\\
\hline
1 & 1 & ${\bf R}_1$ & $J_1$ & $\eta_{1\alpha}$
\\
1 & 1 & ${\bf R}_2$ & $J_1$ & $\eta_{1\beta}$
\\
1 & 1 & $-{\bf R}_1-{\bf R}_2$ & $J_1$ & $\eta_{1\beta}$
\\
\hline
2 & 2 & ${\bf R}_1$ & $J_1$ & $\eta_{1\alpha}$
\\
2 & 2 & ${\bf R}_2$ & $J_1$ & $\eta_{1\beta}$
\\
2 & 2 & $-{\bf R}_1-{\bf R}_2$ & $J_1$ & $\eta_{1\beta}$
\\
\hline
1 & 2 & ${\bf 0}$ & $J_2$ & $\eta_{2\alpha}$
\\
1 & 2 & ${\bf R}_1+{\bf R}_2$ & $J_2$ & $(-1)^m \cdot \eta_{2\beta}$
\\
1 & 2 & ${\bf R}_2$ & $J_2$ & $(-1)^m \cdot \eta_{2\beta}$
\\
\hline
1 & 2 & $-{\bf R}_3$ & $J_2$ & $\eta_{2\alpha}$
\\
1 & 2 & ${\bf R}_1+{\bf R}_2-{\bf R}_3$ & $J_2$ & $(-1)^m \cdot \eta_{2\beta}$
\\
1 & 2 & ${\bf R}_2-{\bf R}_3$ & $J_2$ & $(-1)^m \cdot \eta_{2\beta}$
\\
\end{tabular}
\end{ruledtabular}
\caption{
 $J_{ll'}({\bf R})$ and $\eta_{ll'}({\bf R})$ in (\ref{eq:Hamiltonian-submatrix})
 for
 the (0,0)-flux and (0,$\pi$)-flux ans\"atze.
 The case of $m=1$ corresponds to the (0,0)-flux ansatz
 and
 $m=0$ does to the (0,$\pi$)-flux ansatz.
 \label{tab:MF-links-1}}
\end{table}

\begin{table}
\begin{ruledtabular}
\begin{tabular}{ccccc}
$l$ & $l'$ & ${\bf R}$ & $J_{ll'}({\bf R})$ & $\eta_{ll'}({\bf R})$
\\
\hline
1 & 1 & ${\bf R}_1$ & $J_1$ & $\eta_{1\alpha}$
\\
1 & 2 & ${\bf 0}$ & $J_1$ & $\eta_{1\beta}$
\\
2 & 1 & $-{\bf R}_1$ & $J_1$ & $\eta_{1\beta}$
\\
2 & 2 & $-{\bf R}_1$ & $J_1$ & $\eta_{1\alpha}$
\\
2 & 1 & $-{\bf R}_1+{\bf R}'_2$ & $J_1$ & $\eta_{1\beta}$
\\
2 & 1 & ${\bf R}'_2$ & $J_1$ & $\eta_{1\beta}$
\\
\hline
3 & 3 & $-{\bf R}_1$ & $J_1$ & $\eta_{1\alpha}$
\\
3 & 4 & ${\bf 0}$ & $J_1$ & $\eta_{1\beta}$
\\
4 & 3 & $-{\bf R}_1$ & $J_1$ & $\eta_{1\beta}$
\\
4 & 2 & ${\bf R}_1$ & $J_1$ & $\eta_{1\alpha}$
\\
4 & 3 & $-{\bf R}_1+{\bf R}'_2$ & $J_1$ & $\eta_{1\beta}$
\\
4 & 3 & ${\bf R}'_2$ & $J_1$ & $\eta_{1\beta}$
\\
\hline
1 & 3 & ${\bf 0}$ & $J_2$ & $\eta_{2\alpha}$
\\
1 & 4 & ${\bf R}_1$ & $J_2$ & $(-1)^m \cdot \eta_{2\beta}$
\\
1 & 4 & ${\bf 0}$ & $J_2$ & $(-1)^m \cdot \eta_{2\beta}$
\\
2 & 4 & ${\bf 0}$ & $J_2$ & $\eta_{2\alpha}$
\\
3 & 2 & $-{\bf R}'_2$ & $J_2$ & $(-1)^m \cdot \eta_{2\beta}$
\\
2 & 3 & $-{\bf R}_1+{\bf R}'_2$ & $J_2$ & $(-1)^m \cdot \eta_{2\beta}$
\\
\hline
1 & 3 & $-{\bf R}_3$ & $J_2$ & $\eta_{2\alpha}$
\\
1 & 4 & ${\bf R}_1-{\bf R}_3$ & $J_2$ & $(-1)^m \cdot \eta_{2\beta}$
\\
1 & 4 & $-{\bf R}_3$ & $J_2$ & $(-1)^m \cdot \eta_{2\beta}$
\\
2 & 4 & $-{\bf R}_3$ & $J_2$ & $\eta_{2\alpha}$
\\
3 & 2 & $-{\bf R}'_2+{\bf R}_3$ & $J_2$ & $(-1)^m \cdot \eta_{2\beta}$
\\
2 & 3 & $-{\bf R}_1+{\bf R}'_2-{\bf R}_3$ & $J_2$ & $(-1)^m \cdot \eta_{2\beta}$
\\
\end{tabular}
\end{ruledtabular}
\caption{
 $J_{ll'}({\bf R})$ and $\eta_{ll'}({\bf R})$ in (\ref{eq:Hamiltonian-submatrix})
 for
 the ($\pi$,0)-flux and ($\pi$,$\pi$)-flux ans\"atze.
 The case of $m=0$ corresponds to the ($\pi$,0)-flux ansatz
 and
 $m=1$ does to the ($\pi$,$\pi$)-flux ansatz.
 \label{tab:MF-links-2}}
\end{table}

The mean-field Hamiltonian (\ref{eq:H_MF_momentum_space}) is diagonalized via the Bogoliubov transformation:\cite{BlaizotRipka}
\begin{eqnarray}
 \Psi ({\bf k}) = {\bf M}({\bf k}) \Gamma({\bf k}),
 \nonumber
\end{eqnarray}
where
\begin{eqnarray}
 \Gamma ({\bf k}) 
 = 
 ( \gamma_{1 \uparrow} ({\bf k}), 
   \cdots,
   \gamma_{n_s \uparrow} ({\bf k}), 
   \gamma_{1 \downarrow}^{\dagger} (-{\bf k}), 
   \cdots,
   \gamma_{n_s \downarrow}^{\dagger} (-{\bf k})
 )^T.
 \nonumber
\end{eqnarray}
$\gamma_{l \alpha} ({\bf k})$ satisfies the bosonic statistics,
which imposes a condition on the transformation matrix ${\bf M}({\bf k})$:
\begin{eqnarray}
 {\bf M}({\bf k}) \mathbb{I}_B {\bf M}^{\dagger}({\bf k}) = \mathbb{I}_B,
 \label{eq:eigenvec normalization}
\end{eqnarray}
where
\begin{eqnarray}
 \mathbb{I}_B=
 \left(
 \begin{array}{c|c}
  \mathbb{I}_{n_s} & 0
  \\
  \hline
  0 & -\mathbb{I}_{n_s}
 \end{array}
 \right).
 \nonumber
\end{eqnarray}
Due to this constraint,
$\mathbb{I}_B {\bf D}({\bf k})$ is diagonalized instead of ${\bf D}({\bf k})$,
so the eigenvalue problem has the following form:
\begin{eqnarray}
 {\bf M}^{-1}({\bf k}) \mathbb{I}_B {\bf D}({\bf k}) {\bf M}({\bf k}) = \mathbb{I}_B {\boldsymbol \Omega}({\bf k}),
 \nonumber
\end{eqnarray}
where ${\boldsymbol \Omega}({\bf k})$ is the diagonal matrix of eigenvalues,
\begin{eqnarray}
 {\boldsymbol \Omega}({\bf k}) 
 =
 \textup{diag} 
 ( 
 \omega_{1\uparrow} ({\bf k}), 
 \cdots,
 \omega_{n_s\uparrow} ({\bf k}), 
 \omega_{1\downarrow} (-{\bf k}), 
 \cdots,
 \omega_{n_s\downarrow} (-{\bf k})
 ),
 \nonumber
\end{eqnarray}
In fact, 
$\omega_{l\uparrow} ({\bf k}) = \omega_{l\downarrow} (-{\bf k}) \equiv \omega_{l} ({\bf k})$ ($l=1,\cdots,n_s$),
due to the time reversal symmetry of the mean-field Hamiltonian.
In addition, 
$\omega_{l} ({\bf k})=\omega_{l} (-{\bf k})$
since ${\bf D}(-{\bf k}) = {\bf D}^*({\bf k})$.
The eigenvectors are contained in the columns of ${\bf M}({\bf k})$ 
and normalized according to (\ref{eq:eigenvec normalization}).
By the above transformation,
\begin{eqnarray}
 H_{MF}
 &=&
 N_{uc} \cdot n_s \cdot \epsilon_{gr}
 + 
 \sum_{\bf k} \sum_{l=1}^{n_s} 
 \omega_{l} ({\bf k}) \gamma_{l \alpha}^{\dagger} ({\bf k}) \gamma_{l \alpha} ({\bf k}),
 \nonumber
\end{eqnarray}
where $\epsilon_{gr}$ is the ground state energy per site;
\begin{eqnarray}
 \epsilon_{gr}
 =
 \epsilon_{0}
 +
 \frac{1}{N_{uc} \cdot n_s} 
 \sum_{\bf k} \sum_{l=1}^{n_s} \omega_{l} ({\bf k}).
\end{eqnarray}
The single-spinon excitation spectrum is given by $\omega_{l} ({\bf k})$.

If the spinon excitation becomes gapless at ${\bf k} = \pm {\bf k}^*$,
then the spinon condensate is considered:
\begin{eqnarray}
 {\bf x}({\bf k}) 
 & = &
 \langle \Psi ({\bf k}) \rangle
 \nonumber\\
 & \equiv &
 \left(
  x_{1\uparrow}({\bf k}),
  \cdots,
  x_{n_s\uparrow}({\bf k}),
  x_{1\downarrow}^{*}(-{\bf k}),
  \cdots,
  x_{n_s\downarrow}^{*}(-{\bf k})
 \right)^T.
 \nonumber
 \label{eq:spinon condensate}
\end{eqnarray}
In the existence of the condensate,
the ground state energy is modified into following form:
\begin{eqnarray}
 \epsilon_{gr}
 =
 \epsilon_{0}
 &+&
 \frac{1}{N_{uc} \cdot n_s} 
 \sum_{{\bf k}\ne\pm{\bf k}^*} \sum_{l=1}^{n_s} \omega_{l} ({\bf k})
 \nonumber\\
 &+&
 \frac{1}{N_{uc} \cdot n_s} 
 \sum_{{\bf k}=\pm{\bf k}^*} {\bf x}^{\dagger} ({\bf k}) {\bf D} ({\bf k}) {\bf x} ({\bf k}).
 \label{eq:e_gr of H_MF} 
\end{eqnarray}
The condensate vector is a zero-energy eigenvector found
from
\begin{subequations}
\begin{eqnarray}
 {\bf D}(\pm{\bf k}^*) {\bf x}(\pm{\bf k}^*) = 0,
 \label{eq:eq for the condensate vector}
\end{eqnarray}
where ${\bf x}(\pm{\bf k}^*)$ is normalized to satisfy
\begin{eqnarray}
 \kappa 
 &=& 
 \sum_{{\bf k}=\pm{\bf k}^*} \sum_{l=1}^{n_s} x_{l\mu}^{*} ({\bf k}) x_{l\mu} ({\bf k})
 \nonumber\\
 &+& 
 \sum_{{\bf k} \ne \pm{\bf k}^*} \sum_{l=1}^{n_s} \langle b_{l\mu}^{\dagger} ({\bf k}) b_{l\mu} ({\bf k}) \rangle.
 \label{eq:eq for the spinon density}
\end{eqnarray}
Then, the ground state is determined by solving
\begin{eqnarray}
 \frac{\partial \epsilon_{gr}}{\partial \eta} = 0 ~ ( \eta = \eta_{1\alpha}, \cdots , \eta_{2\beta} ),
 \label{eq:eq for the energy minimization}
\end{eqnarray}
together with (\ref{eq:eq for the condensate vector}) and (\ref{eq:eq for the spinon density}).
\label{eq:self_consistent_equations in k-space}
\end{subequations}
(\ref{eq:self_consistent_equations in k-space}) is the momentum space version of the self-consistent mean-field equations (\ref{eq:self_consistent_equations}).

\section{Mean-field phase diagrams of $(\pi,0)$-flux and $(\pi,\pi)$-flux ans\"atze\label{appendix:(pi,zero/pi)_solution}}

\begin{figure}
 \centering
 \includegraphics[width=0.8\linewidth]{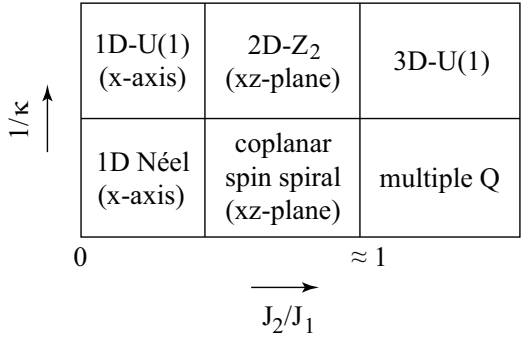}
 \caption{
 Schematic phase diagram of the $(\pi,0)$-flux and $(\pi,\pi)$-flux ans\"atze.
 \label{fig:phase_diagram_pi_zero_ansatz}
 }
\end{figure}

In this appendix, we present the results of the mean-field theories for the $(\pi,0)$-flux and $(\pi,\pi)$-flux ans\"atze.
As will be shown below,
the ans\"atze are energetically less {favoured (within SBMFT)} than the (0,0)-flux and (0,$\pi$)-flux ans\"atze.
Moreover, the semi-classical limit of the $(\pi,0)$-flux and $(\pi,\pi)$-flux ans\"atze 
does not recover the classical spin states found in Sec. \ref{sec:semi-classical-approach}.

\subsection{$(\pi,0)$-flux ansatz}

\begin{figure}
 \centering
 \includegraphics[angle=270,width=0.9\linewidth]{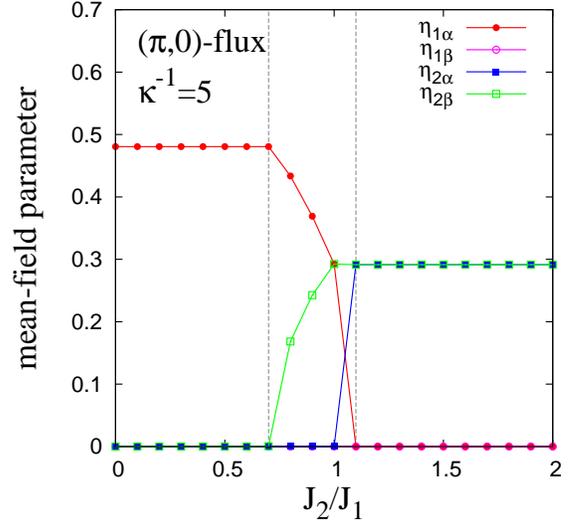}
 \caption{
 (Color online)
 Mean-field parameters of the $(\pi,0)$-flux spin liquid states at $\kappa^{-1}=5$.
 In the plot, $\eta_{2\alpha}$ is all the way zero.
 \label{fig:pi_0-flux_at_kappa_inverse_5}
 }
\end{figure}

\begin{figure*}
 \centering
 \includegraphics[angle=270,width=0.27\linewidth]{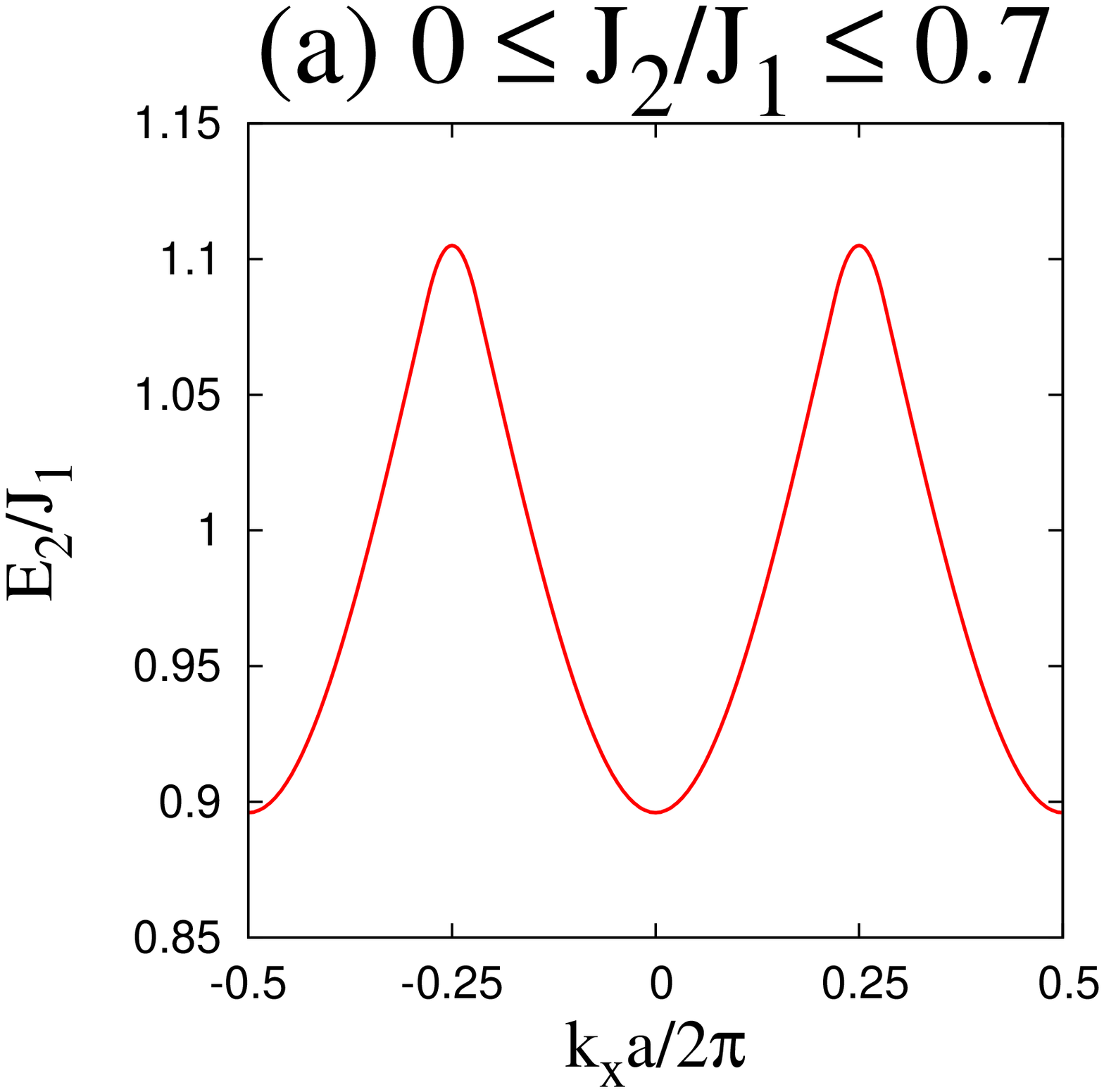}
 \includegraphics[bb=80 170 554 670,clip,width=0.3\linewidth,angle=270]{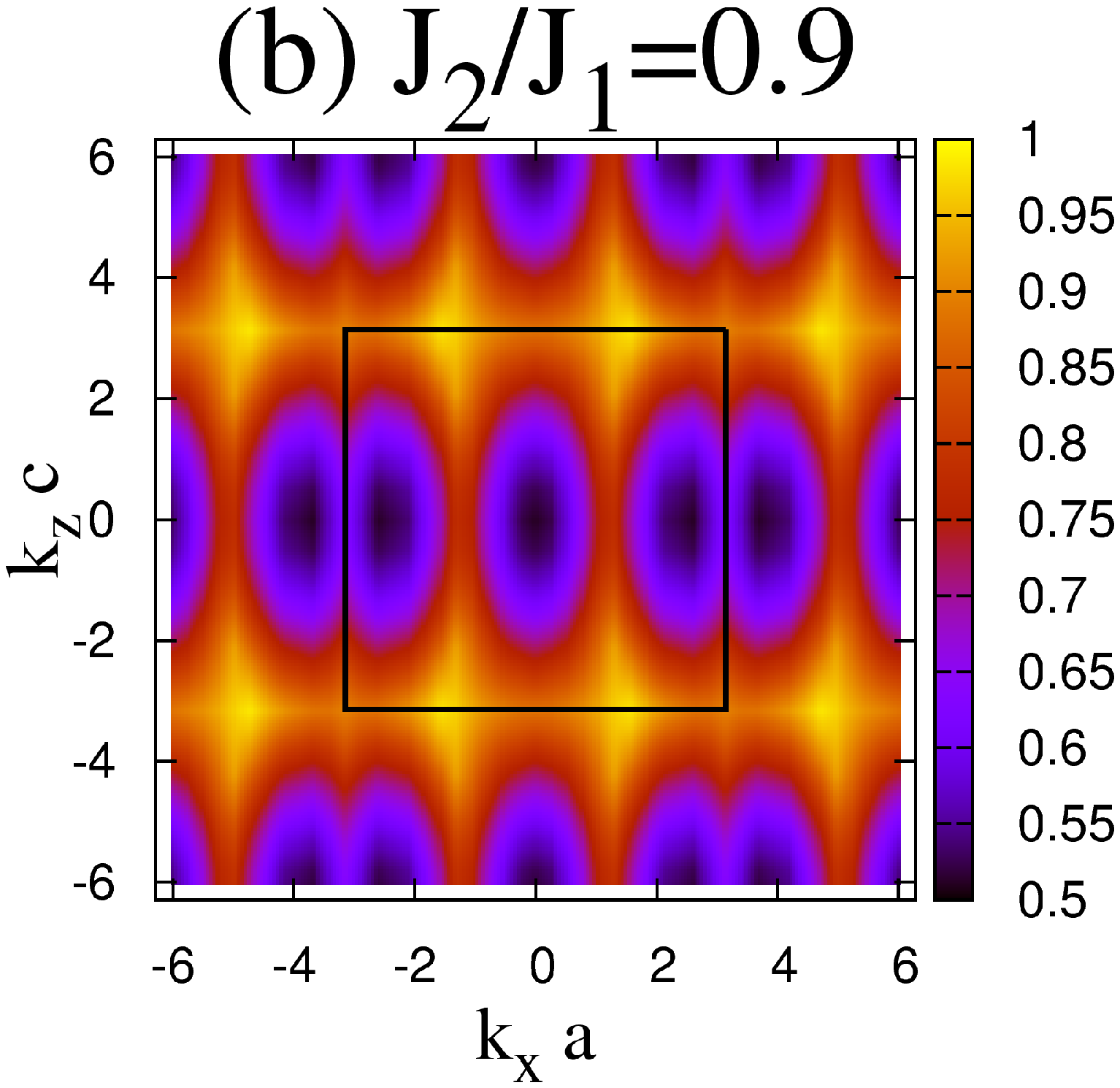}
 \includegraphics[bb=80 170 554 670,clip,width=0.3\linewidth,angle=270]{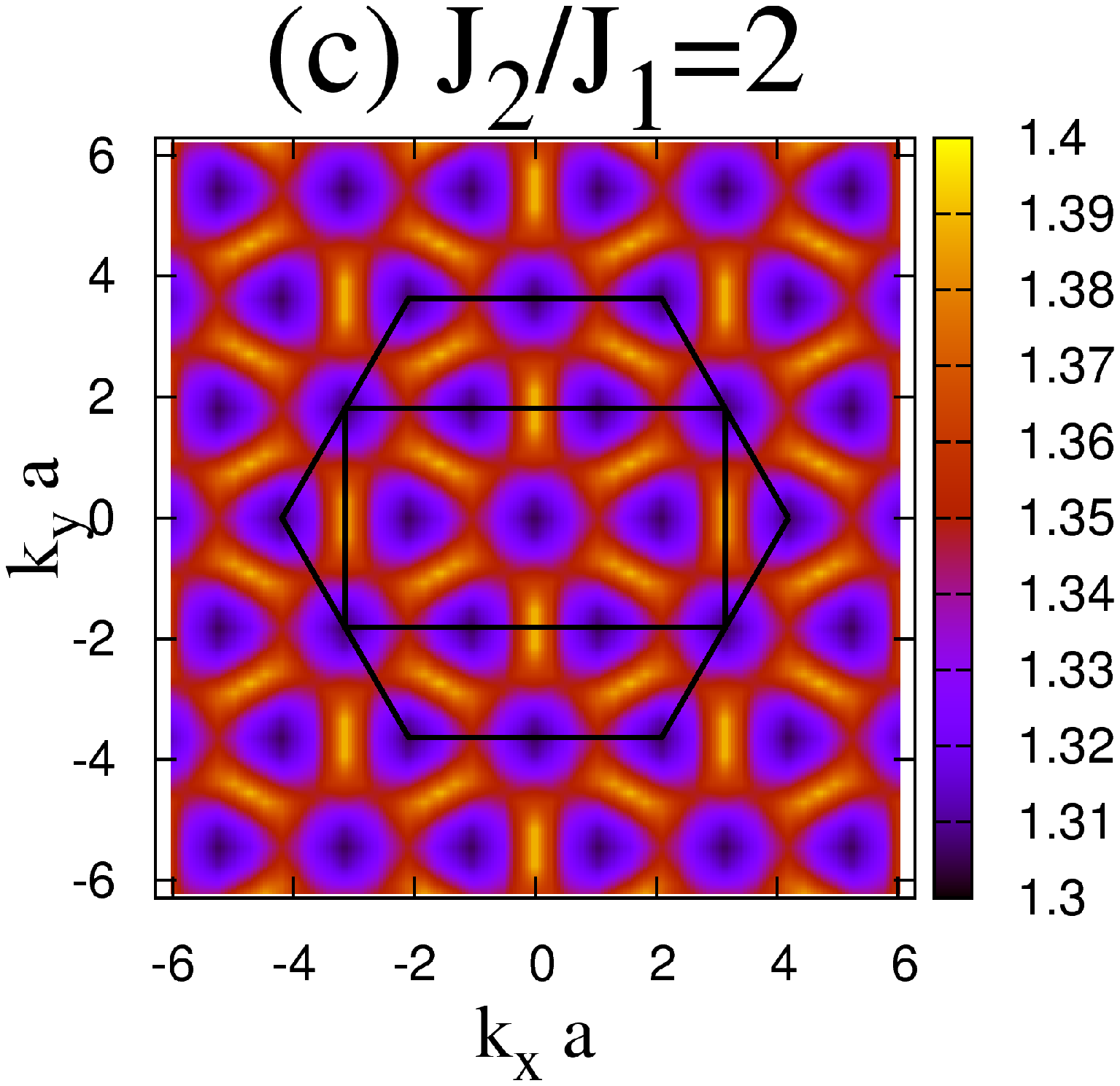} 
 \caption{
 (Color online)
  The lower edge of the two-spinon excitations, $E_2({\bf k})/J_1$, of the ($\pi$,0)-flux spin liquid states at $\kappa^{-1}=5$.
 $E_2({\bf k})$ is defined in (\ref{eq:two-spinon excitation}).
 (a) The $1D$-$U(1)$ spin liquid along the $x$-axis.
 (b) The $2D$-$Z_2$ spin liquid along the $xz$-plane.
 The rectangle denotes the first Brillouin zone of the ansatz in the $k_xk_z$-plane.
 (c) The $3D$-$U(1)$ spin liquid.
 In this case, 
 we plot $E_2({\bf k})$ in the plane $k_z=0$ where the minima of the two-spinon excitation occur.
 The hexagon denotes the first Brillouin zone of the 6H-B lattice structure
 and the rectangle does the Brillouin zone of the ansatz.
 \label{fig:two-spionon-pi_0-flux}
 }
\end{figure*}

Figure \ref{fig:phase_diagram_pi_zero_ansatz} shows a schematic phase diagram for the $(\pi,0)$-flux ansatz.
In the quantum limit ($\kappa^{-1} \gg 1$),
there are three spin liquid phases: 
the $1D$-$U(1)$, $2D$-$Z_2$, and $3D$-$U(1)$ states.
In the semi-classical limit ($\kappa^{-1} \ll 1$),
we find three different kinds of long-range orders:
the 1D N\'eel, spin spiral, and multiple $\bf Q$ states. {The detailed analysis of the phase boundaries is complicated due to the presence of multiple minima in the spinon band structure and hence we only discuss the schematic phase diagram.}
For the spin liquid states, 
the mean-field parameters are depicted in Fig. \ref{fig:pi_0-flux_at_kappa_inverse_5} for $\kappa^{-1}=5$.
The {lower bound} two-spinon dispersion for each spin liquid state is plotted in Fig. \ref{fig:two-spionon-pi_0-flux}.

\begin{figure}
 \centering
 \includegraphics[angle=270,width=0.9\linewidth]{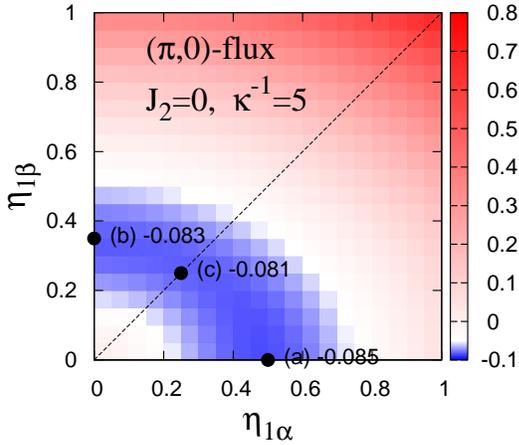}
 \caption{
 (Color online)
 Plot of energy ($\epsilon_{gr}/J_1$) as a function of $\{ \eta_{1\alpha},\eta_{1\beta} \}$ 
 for the $(\pi,0)$-flux ansatz at $J_2=0$ and $\kappa^{-1}=5$.
 Three dots denote the local extrema (the figure indicates the energy at each point).
 (a) The energy minimum point: the $1D$-$U(1)$ spin liquid.
 (b) A local minimum point.
 (c) A saddle point lying on the line $\eta_{1\alpha}=\eta_{1\beta}$: 
 the $\pi$-flux spin liquid state in the 2D triangular lattice.\cite{2006_wang}
 \label{fig:e_gr_of_pi_o_flux}
 }
\end{figure}

\paragraph{\underline{$1D$-$U(1)$ Spin Liquid}}
In the limit of small $J_2/J_1$, only $\eta_{1\alpha}$ is nonzero and the other parameters are all zero.
This leads to an one-dimensional spin liquid along the $x$ axis or $r_1$ direction.
The minimum of the single-spinon gap occurs at the points $(\pm\frac{\pi}{2a},0,0)$. 
Particularly, the 1D spin liquid is stabilized even at the point $J_2=0$, where the system is reduced to {stacked and decoupled} 2D triangular lattices.{This is in contrast to the natural expectation of recovering} the $\pi$-flux state\cite{2006_wang} in the 2D triangular lattice {limit, just like we} obtained the 0-flux state of the triangular lattice {in the $J_2=0$ limit of} the (0,0)-flux and (0,$\pi$)-flux ans\"atze in Sec. \ref{sec:mean-field-phase-diagram}.
{In the present case, we however, find} that the $\pi$-flux state ($\eta_{1\alpha}=\eta_{1\beta}$) is not the {global} energy minimum  
in the space of the parameters $\{ \eta_{1\alpha}$, $\eta_{1\beta} \}$ (see in Fig. \ref{fig:e_gr_of_pi_o_flux}).
{We think that this $1D-U(1)$ spin liquid is an artifact of our mean field theory and it is unstable to other phases. Based on previous analysis\cite{2006_wang}, the $\pi$-flux state appears to be a good candidate phase in this regime. However, a more systematic treatment of this parameter regime is required.}

\paragraph{\underline{$2D$-$Z_2$ Spin Liquid}}
Upon increasing $J_2/J_1$ further, $\eta_{2\beta}$ becomes nonzero in addition to $\eta_{1\alpha}$.
However, $\eta_{1\beta}$ and $\eta_{2\alpha}$ remain zero.
Considering the links of $\eta_{1\alpha}$ and $\eta_{2\beta}$,
one can find that the links have the 2D anisotropic triangular lattice structure along the $xz$-plane or $r_1r_3$-plane.
This yields a $2D$-$Z_2$ spin liquid state.
This spin liquid state is gauge-equivalent to the 0-flux state of the anisotropic triangular lattice.\cite{2001_chung,2006_wang}
In particular,
when $J_2/J_1=1$,
the 0-flux spin liquid state of the isotropic triangular lattice is obtained
(see $\eta_{1\alpha}=\eta_{2\beta}$ at $J_2/J_1=1$ in Fig. \ref{fig:pi_0-flux_at_kappa_inverse_5}).
The $2D$-$Z_2$ spin liquid state has the lowest single-spinon excitations at ($\pm$q,0,0).

\paragraph{\underline{$3D$-$U(1)$ Spin Liquid}}
In the limit of large $J_2/J_1$, $\eta_{1\alpha}$ vanishes and $\eta_{2\alpha}$ has the same magnitude as $\eta_{2\beta}$.
$\eta_{1\beta}$ still remains zero.
The mean-field links of $\{ \eta_{2\alpha}, \eta_{2\beta} \}$ have a three-dimensional, honeycomb-like bipatite lattice structure.
The resulting state is a $3D$-$U(1)$ spin liquid state.
The lowest spinon excitations appear at the four points: 
$\pm(\frac{1}{12}{\bf G}_1 + \frac{1}{4}{\bf G}_2)$ and $\pm(\frac{5}{12}{\bf G}_1 + \frac{1}{4}{\bf G}_2)$,
where ${\bf G}_1=(\frac{2\pi}{a},0,0)$, ${\bf G}_2=(0,\frac{2\pi}{\sqrt{3}a},0)$, and ${\bf G}_3=(0,0,\frac{2\pi}{c})$.
$\{ {\bf G}_n \}_{n=1}^{3}$ are the reciprocal lattice vectors of the lattice vectors $\{ {\bf R}_1, {\bf R}_2^{'}, {\bf R}_3 \}$.


\paragraph{\underline{Long-range orders}}
In large $\kappa$ limit,
spinons are condensed at the minima of the spinon dispersion and a magnetic long-range order is developed.
We find three kinds of magnetic orders.
(1) \emph{1D N\'eel order}:
when $J_2/J_1 \ll 1$, the one-dimensional N\'eel state is stabilized with the ordering wave vector $\pm(\frac{\pi}{a},0,0)$.
(2) \emph{Coplanar spin spiral order}:
on increasing $J_2/J_1$, 
neighboring 1D N\'eel spin chains are correlated along the $xz$-plane and 
the coplanar spin spiral order is developed along the plane.
The ordering wave vectors are given by $\pm(Q,0,0)$ ($Q < \frac{\pi}{a}$).
(3) \emph{Multiple-{\bf Q} state}:
when $J_2/J_1 > 1$, we find magnetic state with multiple ordering wave vectors ({\bf Q}).
The multiple ordering wave vectors result from the spinon condensation at the four points, $\pm(\frac{1}{12}{\bf G}_1 + \frac{1}{4}{\bf G}_2)$ and $\pm(\frac{5}{12}{\bf G}_1 + \frac{1}{4}{\bf G}_2)$, in the $3D$-$U(1)$ spin liquid.
The wave vectors can be identified with the positions of the two-spinon dispersion minima in Fig. \ref{fig:two-spionon-pi_0-flux} (c). {The details of the complicated magnetic structure are presently not understood.}

\subsection{$(\pi,\pi)$-flux ansatz}

The overall phase diagram of the $(\pi,\pi)$-flux ansatz has the same structure as that of the $(\pi,0)$-flux ansatz.
We find a difference between the two ans\"atze in the region of the multiple {\bf Q} state:
they have gauge-inequivalent mean-field states.
In the other regions, their mean-field states are gauge-equivalent.

\subsection{Mean-field Energy comparison}

\begin{figure}
 \centering
 \includegraphics[angle=270,width=0.9\linewidth]{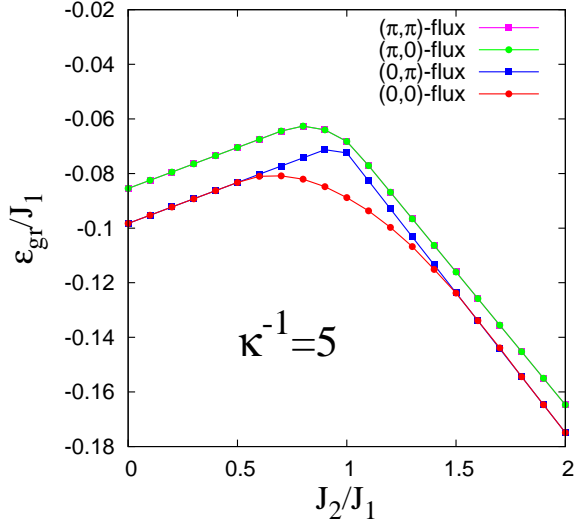}
 \caption{
 (Color online)
 Energy comparison among the ans\"atze with the symmetry group SG$_1$.
 The (0,0)-flux ansatz is the most stable among the ans\"atze with the symmetry group SG$_1$.
 The ($\pi$,0)-flux and ($\pi$,$\pi$)-flux ans\"atze have the same energy 
 since they have gauge-equivalent mean-field states.
 \label{fig:four_ansatze_energy_comparison}
 }
\end{figure}

We close this appendix with an energy comparison (obtained from our SBMFT) among the ans\"atze of SG$_1$.
Fig. \ref{fig:four_ansatze_energy_comparison} shows the ground state energy of each ansatz at $\kappa^{-1}=5$.
The ($\pi$,0)-flux and ($\pi$,$\pi$)-flux ans\"atze have higher energies than the (0,0)-flux ansatz,
which is the most stable among the ans\"atze with the symmetry group SG$_1$.
We find that this pattern is true for general $\kappa^{-1}$.
In the figure, the ($\pi$,0)-flux and ($\pi$,$\pi$)-flux ans\"atze have the same energies
since their mean-field states are gauge-equivalent.

\end{document}